\documentclass[a4paper,12pt]{article}
\def\al{\alpha}
\def\be{\beta}
\def\ga{\gamma} \def\Ga{\Gamma}
\def\ep{\epsilon}
\def\lam{\lambda}
\def\Lam{\Lambda}

\def\calA{{\cal A}}

 \def\calN{{\cal N}} \def\calO{{\cal O}}
\def\calP{{\cal P}}  \def\calR{{\cal R}}
\def\calS{{\cal S}}

\def\del        {  \partial  }
\def\half       {  {1\over 2}  }

\def\ie         {  {\it i.e.}      }

\def\comma          {\, ,}
\def\period         {\, .}
\def\lsim    {\lower .65ex \hbox{\ $\stackrel{<}{\sim}$\ } }
\def\gsim    {\lower .65ex \hbox{\ $\stackrel{>}{\sim}$\ } }
\def\com#1#2   { \left[#1, #2\right]} 
\def\acom#1#2  {\left\{ #1,#2\right\}}
\def\bra#1     {\langle #1 |}
\def\ket#1     {| #1 \rangle}
 
\def\vecii#1#2      {  \left(\begin{array}{c}#1\\#2\end{array}\right)  }
\def\veciii#1#2#3   {  \left(\begin{array}{c}#1\\#2\\#3\end{array}
                     \right)  }
\def\veciv#1#2#3#4  {  \left(\begin{array}{c}#1\\#2\\#3\\#4
                                 \end{array}\right)  }
\def\vecfv#1#2#3#4#5 {  \left(\begin{array}{c}#1\\#2\\#3\\#4\\#5
                                 \end{array}\right)  }

\def\matrixii#1#2#3#4            {  \left(\begin{array}{cc}#1&#2\\#3&#4
                                       \end{array}\right) }
\def\matrixiii#1#2#3#4#5#6#7#8#9 {  \left(\begin{array}{ccc}#1&#2&#3\\
                                     #4&#5&#6\\#7&#8&#9\end{array}
                               \right)  }
\def\mativ#1#2#3#4               {  \left(\begin{array}{cccc}
                                       #1\\#2\\#3\\#4\end{array}\right) }
\def\matv#1#2#3#4#5              {  \left(\begin{array}{ccccc}
                                     #1\\#2\\#3\\#4\\#5\end{array}
                              \right)  }
\def\eqabegin         {  \begin{eqnarray}  }
\def\eqaend           {  \end{eqnarray}  }
\def\nn               {  \nonumber  }
\def\bracetwo#1#2     {  \left\{ \begin{array}{l} #1 \\ #2 \end{array}
                         \right.  }
\def\bracetwocases#1#2#3#4  {   \left\{ \begin{array}{ll} #1 &
                                 \qquad #2 \\
                                 #3 & \qquad #4 \end{array} \right.  }
\def\bracebegin#1     {  \left\{ \begin{array}{#1}   }
\def\braceend         {  \end{array}\right.   }
\def\parn              {  \par\noindent }

\def\parmedskip        {  \par\medskip  }

\def\parag#1           {\paragraph{#1} \mbox{ }\parmedskip\noindent}
\def\msection#1      {  \begin{center} \section{#1} \end{center}   }
\def\nsection#1      {  \let\boldface\bf \def\bf{} \section{#1}
                           \let\bf\boldface   }
\def\mnsection#1     {  \begin{center} \nsection{#1} \end{center}  }
\def\capsection#1    {  \let\boldface\bf \def\bf{\sc} \section{#1}
                           \let\bf\boldface   }
\def\mcapsection#1   {  \begin{center} \capsection{#1} \end{center} }
\newcommand{\nullify}[1]{}
\def\papertitlepage{\baselineskip 3.5ex \thispagestyle{empty}}
\def\Title#1{\baselineskip 1cm \vspace{1.5cm}\begin{center}
 {\Large\bf #1} \end{center} 
\vspace{0.5cm}}
\def\Authors#1{\begin{center} {\it #1} \end{center}}
\def\Abstract{\vspace{1.0cm}\begin{center} {\large\bf Abstract} 
           \end{center} \par\bigskip}
\def\Komabanumber#1#2#3{\hfill \begin{minipage}{4.2cm} UT-Komaba #1
              \parn #2 
              \parn #3 \end{minipage}}
\renewcommand{\thefootnote}{\fnsymbol{footnote}}
\renewenvironment{thebibliography}{\pagebreak[3]\par\vspace{0.6em}
\begin{flushleft}{\large \bf References}\end{flushleft}
\vspace{-1.0em}

\begin{enumerate}\if@twocolumn\baselineskip=0.6em\itemsep -0.2em
\else\itemsep -0.2em\fi\labelsep 0.1em}{\end{enumerate}}
\usepackage{graphicx}
\usepackage{latexsym}
\usepackage{amsmath}
\usepackage{amssymb}
\renewenvironment{thebibliography}{\pagebreak[3]\par\vspace{0.6em}
\begin{flushleft}{\large \bf References}\end{flushleft}
\vspace{-1.0em}

\begin{enumerate}\if@twocolumn\baselineskip=0.6em\itemsep -0.2em
\else\itemsep -0.2em\fi\labelsep 0.1em}{\end{enumerate} }
\def\Ntil{\tilde{N}}

\def\Ktil{{\tilde{K}}}

\def\Qtil{{\tilde{Q}}}

\def\Ltil{\tilde{L}}
\def\Stil{\tilde{S}}
\def\Ptil{{\tilde{P}}}
\def\qtil{{\tilde{q}}}
\def\ftil{\tilde{f}}
\def\gtil{\tilde{g}}
\def\Etil{\tilde{E}}
\def\Gabar{\bar{\Ga}}

\def\Qbar{\bar{Q}}

\def\Qhat{\hat{Q}}
\def\Khat{\hat{K}}
\def\Nhat{\hat{N}}
\def\gcom#1#2{{\left[ #1, #2\right\}}}
\def\bk#1#2{{\langle #1 | #2 \rangle }} % bracket
\def\calNbar{\bar{\calN} }

\def\lampinv{\lam_+^{-1}}
\def\lamp{\lam_+}

\def\delx{\del x}

\def\Llam{L_{(\lam)}}

\def\QB{Q_B}

\def\Xdot{\dot{X}}
\def\Ach{\check{A}}

\def\adot{{\dot{a}}}
\def\bdot{{\dot{b}}}
\def\fivetil{{\tilde{5}}}
\def\ub#1{\underbrace{#1}}
\def\ptil{{\tilde{p}}}
\def\deltac{\delta_{c}}
\def\ovsqtwo{{1\over \sqrt{2}}}

\def\lambar{\bar{\lambda}}
\def\chibar{\bar{\chi}}
\def\npb#1{Nucl. Phys. {\bf B#1}}
\def\prd#1{Phys. Rev. {\bf D#1}}
\def\jhep#1{JHEP {\bf #1}}
\def\hepth#1{ hep-th/#1}
\def\plb#1{Phys. Lett. {\bf B#1}}
\begin{document}
%\nullify{
%%%%%%%%%%%%%%%%%%%%%%%%%%%%
\papertitlepage
\vspace*{0cm}
\Komabanumber{04-5}{hep-th/0404141} {April, 2004}
%%%%%%%%%%%%%%%%%%%%%%%%%%%%
\Title{Relating Green-Schwarz
 and Extended Pure Spinor Formalisms
by Similarity Transformation} 
%\vspace{1cm}
\Authors{{\sc Yuri Aisaka\footnote[2]{yuri@hep1.c.u-tokyo.ac.jp} 
 and Yoichi Kazama
\footnote[3]{kazama@hep1.c.u-tokyo.ac.jp}
\\ }
\vskip 3ex
 Institute of Physics, University of Tokyo, \\
 Komaba, Meguro-ku, Tokyo 153-8902 Japan \\
  }
%\vspace{0.7cm}
\baselineskip .7cm

%%%%%%%%%%%%%%%%%%%%%%%%%%%%%%%%%%%%%%%%%(*.*)
\numberwithin{equation}{section}
\numberwithin{figure}{section}
\numberwithin{table}{section}
%%%%%%%%%%%%%%%%%%%%%%%%%%%%%%%%%%%%%%%%%

%%%%%%%%%%%%%%%%%%%%%%%%%%%%%%%%%%%%%%%
\parskip=0.9ex
%%%%%%%%%%%%%%%%%%%%%%%%%%%%%%%%%%%%%%%

%%%%%%%%%%%%%%%%%%%%%%%%%%%%%%%%%%%%%%%
\Abstract

 In order to gain deeper understanding of  pure-spinor-based 
 formalisms of superstring, 
an explicit  similarity transformation is constructed 
 which provides operator mapping between 
 the light-cone Green-Schwarz (LCGS) formalism and the 
extended pure spinor (EPS) formalism, a recently proposed 
generalization of the Berkovits' formalism in an enlarged space. 
By applying a systematic procedure developed in our previous 
work, we first construct an analogous mapping in the bosonic string 
relating the BRST and the light-cone formulations. This provides 
sufficient insights and allows us to construct the desired 
 mapping in the  more intricate case of superstring as well.
The success of the construction owes much to the enlarged field space
 where pure spinor constraints are removed and to the existence 
 of the ``$B$-ghost'' in the EPS formalism.

\newpage
%%%%%%%%%%%%%%%%%%%%%%
\baselineskip 3.5ex
%%%%%%%%%%%%%%%%%%%%%%%%%%%%%%%%%%%%%%%%%%%%%%%%%%%%%%%%%%%%%%%%
\section{Introduction}  
%%%%%%%%%%%%%%%%%%%%%%%%%%%%%%%%%%%%%%%%%%%%%%%%%%%%%%%%%%%%%%%%
% For any sentence starting with ``From'', write {}From 
\renewcommand{\thefootnote}{\arabic{footnote}}
%{\sc ! Reorganize references. Add recent papers by Stony Brook group.} 
%\parbigskipn
%%%%%%%%%%%%%%%%%
Quantization of superstring in which  both the Lorentz symmetry and 
the supersymmetry are manifest is a long standing problem of 
 prime importance. Not long ago, a promising new approach based on 
 the concept of pure spinor (PS) \cite{Cartan}-\cite{Howe91}
 was put forward  by Berkovits \cite{Berk0001}, after many 
 partially successful attempts of various kinds
 \cite{Siegel86}\nocite{BVparticles,BVparticlesII,%
HSstring,TWstring,Hybrid}-\cite{U5form}. In this formulation 
  the physical states of superstring are obtained as the cohomology 
 of a BRST-like operator $Q_B=\int[dz] \lam^\al d_\al$, where 
$d_\al$ is the spinor covariant derivative and $\lam^\al$ is a bosonic 
 chiral pure spinor satisfying the non-linear constraints 
$\lam^\al \ga^\mu_{\al\be} \lam^\be =0$, which render $Q_B$  nilpotent. 
All the independent worldsheet fields in this formalism are taken to be
  free and form a conformal field theory with vanishing center. This 
 feature allows one to construct $Q_B$-invariant vertex operators explicitly 
and certain covariant rules are proposed to compute the 
 scattering amplitudes, which reproduce  known results
\cite{Berk0001}\cite{Berk0004}\nocite{Berk0104}-\cite{Trivedi:2002mf}. 
We refer the reader to \cite{Berk0006}\nocite{%
Berk0006,Berkovits:2000wm,Berk0009,Berk0104,Berk01050,Berk01051,Berk0112,%
Berk0201,Berk0203,Berk0204,Berk0205,Oda0109,Matone0206,stonybrook,
Chesterman,AK1,AK2}-\cite{Oda0302}
 for further developments 
 and \cite{Berk0209} for a review up to a certain point.

Although highly compelling evidence has already 
been accumulated in support of the 
 basic idea of this  formulation, there are many points yet 
to be clarified and 
 improved. Among them the most fundamental are the questions of the underlying 
 action, its symmetries and how it should be quantized covariantly. 
In this regard it is ironic that the very concept of pure spinor 
with its characteristic quadratic constraints appears to cause 
some  problems. 
Since only part of its components are independent free fields,
 Berkovits' formulation, in strict sense,  is only $U(5)$-covariant.
More importantly, the constrained Hilbert space is not large enough so that 
 proper hermitian inner product is hard to define \cite{AK1}, the ``$B$-ghost''
 needed to generate the energy-momentum tensor $T(z)$ via 
 $\acom{Q_B}{B(z)} = T(z)$ is difficult to construct, etc. 
 After all  such non-linear constraints would have to be avoided 
 in quantization of the underlying  action. 
 
This motivated several groups to try to remove the PS constraints by 
  introducing  some
 finite number of compensating ghosts. One approach, described 
 in a series of papers in \cite{stonybrook},
 is to construct a new nilpotent  BRST-like 
 charge, enforcing covariance at every step, by several methods including 
  one based on the gauged WZW structure. A price for the manifest 
 covariance is that it appears non-trivial to assure the 
 correct cohomology at the present stage. 

An alternative formulation without PS constraints,
 which is in a sense complimentary to  the above, was proposed  
 in our previous work \cite{AK1}.
 In this approach, to be briefly reviewed in Sec.~2, 
  equivalence with the Berkovits' cohomology 
 is manifest by construction whereas the Lorentz symmetry is broken to $U(5)$ 
in the newly added ghost sector. In this sense it is 
a minimal extension of the Berkovits' theory. The enlargement of the 
 Hilbert space, however, has a definite advantage. An appropriate ``$B$-ghost''
 field was obtained in a simple form using the added ghosts
 \cite{AK1}, and in a subsequent work \cite{AK2}, 
we have been able to construct a quantum similarity transformation 
which connects the BRST operator for the conventional RNS formalism 
 to the one for the EPS without  need of singular operations 
 encountered in a previous attempt in the original PS formalism
 \cite{Berk0104}. 

Although this explicit  demonstration of the equivalence 
 to the RNS via a similarity transformation added further credibility 
 to the EPS formalism, the transformation was rather complicated 
 due to the necessity of conversion between the spacetime vector 
 $\psi^\mu$ in  RNS and the spacetime spinor $\theta^\al$ in EPS, 
which involved mixing of the RNS ghosts as well in an essential way. 
Also, since one has to use the formulation of RNS in the so-called 
``large'' Hilbert space \cite{FMS86}, the degeneracy due to ``pictures''
had to be carefully dealt with \cite{Berk0104}. 

This suggests that it would be of great interest to 
 try to connect the EPS formalism to the light-cone quantized Green-Schwarz
(LCGS) formalism, where none of the above complications exists and 
the physical content of the theory is transparent. In fact at an early stage
 of the development of the PS formalism, such an attempt was made 
 in \cite{Berk0006}. In this work, $SO(8)$ parametrization of the 
 PS conditions is used, which turned out to be infinitely reducible. 
As a consequence, indefinite number of ghosts for ghosts had to be
 introduced and the demonstration of the equivalence with the light-cone
 formulation was quite cumbersome even conceptually.
 
In the present work, we shall make use of  the systematic method 
developed in \cite{AK2} to relate the LCGS formalism 
 to the EPS formalism by constructing an appropriate  similarity
 transformation. There are several advantages for making the 
 connection by a similarity transformation. The function of such a 
 transformation is to reorganize the Hilbert space into the direct 
 product of physical and unphysical sectors $\ket{{\rm phys}} \otimes 
\ket{{\rm unphys}} $,  thereby effectively achieving the gauge fixing 
without discarding any degrees of freedom (DOF). In other words, it emphasizes 
 the aspect of decoupling of the unphysical DOF rather than their elimination. 
 Moreover, in this process all the 
operatorial relations, in particular the (anti-)commutation relations 
  and the symmetry structure, are preserved. We expect that the understanding
 gained by such a construction should prove quite useful for 
future attempts to uncover the underlying action and its symmetries. 

Let us describe more explicitly the outline of our work. 
 We begin  with a brief review of the PS and the EPS formalisms in  Sec.~2.
 Then in Sec.~3 we consider, as a warm up, the bosonic string and construct a similarity transformation
 which provides a mapping between 
 the BRST  and  the light-cone formalisms. 
Besides proving the equivalence of two formulations in a manifest manner, 
 this transformation explicitly converts the transverse oscillators $\al^i_n$ 
 of the light-cone formalism into the DDF spectrum generating oscillators 
\cite{DDF} 
 $A^i_n$ of the BRST formalism. 
Making use of the understanding gained by this exercise, we attack in 
 Sec.~4 our main problem of relating the LCGS and the EPS 
 formalisms for superstring by a similarity mapping. This will be 
 divided into three stages. In Sec.~4.1, starting from the BRST version of
 the light-cone formalism we make two simple similarity transformations 
 to bring the ``light-cone BRST operator'' $\Qbar$ 
into a form called $\Qtil$, 
 in which all the  fields needed in EPS formalism are visible. Then in 
 Sec.~4.2, we develop further similarity transformations which simplify 
 the BRST operator $\Qhat$ for the EPS formalism into $\Qtil$. 
Finally in Sec.~4.3 we show that in a suitably defined Hilbert space 
 the cohomology of $\Qbar$ indeed gives the light-cone spectrum. 
We summarize our results and discuss future problems in Sec.~5. 
Two appendices are provided to give some details not covered in the main 
 text: In Appendix A our conventions and some useful formulas are collected. 
Appendix B gives further details of the construction 
 of the similarity transformations in the bosonic and the superstring cases. 

%%%%%%%%%%%%%%%%%
\section{A Brief Review of PS and EPS Formalisms  }
%%%%%%%%%%%%%%%%%
To make this article reasonably self-contained and 
 to explain our notations, let us begin 
 with a brief review of the essential features of the PS and the EPS
formalisms.  
%%%%%%%%%%%%
\subsection{PS Formalism}
%%%%%%%%%%%%%%
The central idea of the pure spinor formalism 
\cite{Berk0001} is that the physical states of superstring 
can be described as the elements of the cohomology of a BRST-like 
 operator $\QB$ given by\footnote{For simplicity we will use 
 the notation $[dz] \equiv dz/(2\pi i)$ throughout.}
\begin{eqnarray}
\QB &=& \int [dz] \lam^\al(z) d_\al(z) \comma \label{defQ}
\end{eqnarray}
where $\lam^\al$ is a $16$-component bosonic chiral spinor 
 satisfying the  pure spinor constraints 
\begin{eqnarray}
\lam^\al \ga^\mu_{\al\be} \lam^\be &=& 0 \comma \label{psconst}
\end{eqnarray}
and $d_\al$ is the spinor covariant derivative given by 
\begin{eqnarray}
d_\al &=& p_\al + i\del x_\mu (\ga^\mu\theta)_\al + \half 
(\ga^\mu\theta)_\al (\theta \ga_\mu \del \theta) \period \label{defdal}
\end{eqnarray}
$x^\mu$ and $\theta^\al$ are, respectively, the basic bosonic and 
ferminonic worldsheet fields describing a superstring, which transform 
 under the spacetime supersymmetry with global spinor parameter $\ep^\al$ 
as $\delta \theta^\al = \ep^\al$, 
$\delta x^\mu = i\ep \ga^\mu \theta$.
$x^\mu$ is self-conjugate and satisfies  $x^\mu(z) x^\nu(w) 
 = -\eta^{\mu\nu} \ln (z-w)$, while $p_\al$ serves as the conjugate 
 to $\theta^\al$ in the manner $\theta^\al(z) p_\be(w) 
 = \delta^\al_\be (z-w)^{-1}$.
 $\theta^\al$ and $p_\al$ carry 
 conformal weights 0 and 1 respectively. With such free field operator 
 product expansions (OPE's), 
$d_\al$ satisfies the following OPE with itself, 
\begin{eqnarray}
d_\al (z) d_\be(w) &= & {2i \ga^\mu_{\al\be} \Pi_\mu(w) 
\over z-w} \comma \label{ddope}
\end{eqnarray}
where $\Pi_\mu$ is the basic superinvariant combination 
\begin{eqnarray}
\Pi_\mu &=& \del x_\mu -i\theta \ga_\mu \del \theta \period 
\label{defPi}
\end{eqnarray}
Then, due to the pure spinor constraints (\ref{psconst}), $\QB$ is easily 
 found to be nilpotent and the  constrained cohomology of $\QB$ 
 can be defined. The basic superinvariants $d_\al, \Pi^\mu$ and 
$\del\theta^\al$  form the closed algebra 
\begin{eqnarray}
d_\al(z) d_\be(w) &=& {2i\ga^\mu_{\al\be} \Pi_\mu (w)\over z-w} 
\comma \label{dd2}\\
d_\al(z) \Pi^\mu(w) &=& {-2i(\ga^\mu \del\theta)_\al(w) \over z-w} 
\comma \label{dPi2}\\
\Pi^\mu(z) \Pi^\nu(w) &=& {-\eta^{\mu\nu} \over (z-w)^2} 
\comma \label{PiPi2}\\
d_\al(z) \del \theta^\be(w) &=& {\delta^\be_\al \over (z-w)^2} 
\comma \label{ddelth2}
\end{eqnarray}
which has  central charges and hence is essentially of {\it second class}. 

Although eventually the rules for computing the scattering 
 amplitudes are formulated in a Lorentz covariant manner, proper
quantization of the pure spinor $\lam$ can only be performed  
by solving the PS constraints (\ref{psconst}), which inevitably 
 breaks covariance in  intermediate steps.
 One convenient scheme is the so-called $U(5)$ formalism\footnote{
Although our treatment applies equally well to $SO(9,1)$ and $SO(10)$ 
 groups, we shall use the terminology appropriate for $SO(10)$,
 which contains $U(5)$ as a subgroup. Some further explanation 
 of our conventions for $U(5)$ parametrization is given in Appendix A.},
 in which 
a chiral and an anti-chiral spinors $\lam^\al$ and $\chi_\al$, respectively, 
are decomposed in the following way
\begin{eqnarray}
\lam^\al &=& (\lam_+, \lam_{PQ}, \lam_\Ptil) \sim  (1,10,\overline{5}) \comma
\label{lamal}  \\
\chi_\al &=& (\chi_-, \chi_{\Ptil\Qtil}, \chi_P) 
\sim (1,\overline{10}, 5)\comma 
 \qquad (P,Q, \Ptil,\Qtil = 1 \sim 5)\comma   \label{chial}
\end{eqnarray}
where we have indicated how they transform under $U(5)$, with a 
 tilde on the $\bar{5}$ indices. On the other hand, an arbitrary
 Lorentz vector 
 $u^\mu$ is split into $5+\overline{5}$ of $U(5)$ as 
\begin{eqnarray}
u^\mu &=& 2 (e^{+\mu}_P u^-_{\Ptil} + e^{-\mu}_\Ptil u^+_P ) \comma 
\label{uU5}
\end{eqnarray}
where the projectors $e^{\pm \mu}_P$, defined by
\begin{eqnarray}
e^{\pm \mu}_P \equiv \half (\delta_{\mu,2P-1} \pm i \delta_{\mu, 2P})\comma 
\label{defeproj}
\end{eqnarray}
 enjoy the properties
\begin{eqnarray}
&& e^{\pm \mu}_P e^{\pm \mu}_Q = 0 \comma \qquad 
  e^{\pm \mu}_P e^{\mp \mu}_Q = \half \delta_{PQ} \comma \\
&&  e^{+\mu}_P e^{-\nu}_\Ptil +e^{-\mu}_\Ptil e^{+\nu}_P
 = \half \delta^{\mu\nu} \period
\end{eqnarray}
Thus, one can invert the relation (\ref{uU5}) in the form
\begin{eqnarray}
u^+_P &=& e^{+\mu}_P u_\mu \comma \qquad u^-_\Ptil = e^{-\mu}_\Ptil u_\mu
\period \label{upm}
\end{eqnarray}
In this scheme the 
 pure spinor constraints reduce to 
 5 independent conditions:
\begin{eqnarray}
\Phi_\Ptil \equiv  \lam_+\lam_\Ptil -{1\over 8}
\ep_{\Ptil \Qtil \tilde{R} \tilde{S} \tilde{T}}\lam_{QR}
 \lam_{ST} =0  \comma  \label{defPhi} 
\end{eqnarray}
and hence $\lam_\Ptil$'s are solved in terms of $\lamp$ and $\lam_{PQ}$. 
Therefore the number of independent components of a pure spinor is 
 11 and together with all the other fields (including the conjugates 
 to the independent components of $\lam$) the entire system constitutes 
 a free CFT with vanishing central charge. 

The fact that the constrained cohomology of $\QB$ is in one to one 
 correspondence with the light-cone degrees of freedom of superstring 
 was shown in \cite{Berk0006} using the $SO(8)$ parametrization of a
 pure spinor. Besides being non-covariant, 
this parametrization is infinitely redundant and an indefinite 
 number of supplementary ghosts had to be introduced. Nonetheless, 
 subsequently the Lorentz invariance of the cohomology 
was demonstrated in \cite{Berk01051}. 

The great advantage of this formalism is that one can compute the 
 scattering amplitudes in a manifestly super-Poincar\'e covariant
 manner. For the massless modes, 
the physical unintegrated vertex operator is given by a simple form 
\begin{eqnarray}
U = \lam^\al A_\al(x,\theta) \comma \label{unintvert} % ZZZ
\end{eqnarray}
where $A_\al$ is a spinor superfield satisfying the ``on-shell'' condition
$(\ga^{\mu_1\mu_2 \ldots \mu_5})^{\al\be} D_\al A_\be = 0$ 
with $D_\al = {\del\over \del\theta^\al} -i(\ga^\mu\theta)_\al
{\del \over \del x^\mu} $. 
Then, with the pure spinor constraints, one easily verifies $\QB U=0$ % ZZZ
and moreover finds that $\delta U =\QB\Lam$ % ZZZ
 represents the gauge transformation
 of $A_\al$. Its integrated counterpart $\int[dz]V(z)$, needed % ZZZ
 for calculation of $n$-point amplitudes with $n\ge 4$, is 
 characterized by $\QB V =\del U$ and was constructed % ZZZ ZZZ
to be of the form \cite{Berk0001,Siegel86}
\begin{eqnarray}
V = \del\theta^\al A_\al + \Pi^\mu B_\mu + d_\al W^\al  % ZZZ
 + \half \Llam^{\mu\nu} F_{\mu\nu} \period \label{intvert}
\end{eqnarray}
Here, $B_\mu =(i/16) \ga^{\al\be}_\mu D_\al A_\be$ is the gauge superfield, 
 $W^\al =(i/20) (\ga^\mu)^{\al\be}(D_\be B_\mu-\del_\mu A_\be)$ is 
 the gaugino superfield,  $F_{\mu\nu}=\del_\mu B_\nu -\del_\nu B_\mu$ 
 is the field strength superfield and $\Llam^{\mu\nu}$ is the Lorentz generator
 for the pure spinor sector. 

With these vertex operators, the scattering amplitude is expressed 
 as 
$\calA =\langle U_1(z_1)U_2(z_2)$ \allowbreak $ U_3(z_3) 
\int [dz_4] V_4(z_4) \cdots \int [dz_N] V_N(z_N) \rangle$,
 and  can be 
 computed in a covariant manner with certain rules assumed for
 the integration over the zero modes of $\lam^\al$ and $\theta^\al$. 
The proposed prescription enjoys a number of required properties and 
 leads to results which agree with those obtained in the RNS formalism
 \cite{Berk0001}\cite{Berk0004}\nocite{Berk0104}-\cite{Trivedi:2002mf}.

%%%%%%%%%%%%%
\subsection{EPS Formalism}
%%%%%%%%%%%%%
Although the PS formalism briefly reviewed above has a number of 
 remarkable features, for the reasons stated in the introduction, 
 it is desirable to remove the PS constraints by extending 
 the field space. Such an extension was achieved in a minimal manner 
 in \cite{AK1}. Skipping all the details, we give below the essence of the 
 formalism. 

Instead of the basic superinvariants forming the essentially 
 second class algebra (\ref{dd2}) 
 $\sim$ (\ref{ddelth2}), we introduce the four types of composite 
 operators
\begin{eqnarray}
j  &=& \lam^\al d_\al \comma \label{defj} \\
\calP_P &=& \calN^\mu_P \Pi_\mu \comma \label{defcalP} \\
\calR_{PQ} &=& 2i\lampinv \calN^\mu_P (\ga_\mu \del \theta)_Q \comma 
\label{defcalR} \\
\calS_{PQ} &=& -(\del \calN^\mu_P) \calN^\mu_Q \comma \label{defcalS} 
\end{eqnarray}
where $\calN^\mu_P$ are a set of five Lorentz vectors which are 
 {\it null},  \ie $\calN^\mu_P \calN^\mu_Q =0$, defined by 
\begin{eqnarray}
\calN^\mu_P \equiv -4(e^{+\mu}_P -\lampinv \lam_{PQ} e^{-\mu}_\Qtil ) 
\period \label{defcalN}
\end{eqnarray}
Note that $j$ is the BRST-like current of Berkovits {\it now without 
 PS constraints}. 
The virtue of this set of operators is that, by using the basic relations 
(\ref{dd2}) $\sim$ (\ref{ddelth2}),
  they can be  shown to form a closed algebra 
 which is of {\it first class}, namely without any central charges. 
This  allows one to build a BRST-like nilpotent charge $\Qhat$ 
associated to this algebra. Introducing five 
sets of ferminonic ghost-anti-ghost pairs $(c_\Ptil, b_P)$ carrying 
 conformal weights $(0,1)$ with 
the OPE
\begin{eqnarray}
 c_\Ptil(z) b_Q(w) = {\delta_{PQ} \over z-w} 
\comma \label{bIcIope}
\end{eqnarray}
and making use of the powerful scheme known as homological perturbation
 theory \cite{HenTb}, $\Qhat$ is constructed as 
\begin{eqnarray}
\Qhat &=& \delta + Q + d_1 + d_2  \comma  \label{Qhat}
\end{eqnarray}
where 
\begin{eqnarray}
\delta &=& -i\int [dz] b_P \Phi_\Ptil \comma \qquad 
Q = \int [dz] j \comma \label{deltaQ}\\
d_1 &=& \int[dz] c_\Ptil \calP_P \comma \qquad 
d_2 = -{i\over 2}\int[dz] c_\Ptil c_\Qtil \calR_{PQ} \period \label{d1d2}
\end{eqnarray}
The operators $(\delta, Q, d_1, d_2)$ carry degrees $(-1,0,1,2)$ 
under the grading  $\deg (c_\Ptil) = 1$, $\deg (b_P) =-1$, 
$\deg (\mbox{rest})=0$ and the nilpotency of $\Qhat$ follows 
 from the first class algebra mentioned above. 

The crucial point of this construction is that by the main theorem of 
 homological perturbation theory the cohomology of $\Qhat$ is guaranteed to 
 be equivalent to that of $Q$ with the constraint $\delta=0$, \ie with 
 $\Phi_\Ptil=0$, which are nothing but the PS constraints (\ref{defPhi}).
 Moreover,  the underlying logic of this proof can be adapted to construct 
the massless vertex operators, both unintegrated and integrated, which 
 are the generalization of the ones shown in 
(\ref{unintvert}) and (\ref{intvert}) for the PS formalism. 
It should also be emphasized that in this formalism, 
 due to the enlarged field space, 
 one can construct the ``$B$-ghost'' field which realizes 
 the important relation  $\acom{\Qhat}{B(z)} = T^{EPS}(z)$.

To conclude this  brief review, let us summarize the basic fields 
of the EPS formalism, their OPE's, the energy-momentum tensor 
$T^{EPS}(z)$ and  the $B$-ghost field. Apart from the $(c_\Ptil, b_P)$ ghosts 
 given in (\ref{bIcIope}), the basic fields are the conjugate pairs 
$(\theta^\al, p_\al)$, $(\lam^\al, \omega_\al)$, both of which carry 
 conformal weights $(0,1)$,  and the string coordinate
 $x^\mu$. Non-vanishing OPE's among them are
\begin{eqnarray}
\theta^\al(z) p_\be(w) &=& {\delta^\al_\be \over z-w}\comma 
 \quad \lam^\al(z) \omega_\be(w) = {\delta^\al_\be \over z-w} 
\comma \quad x^\mu(z) x^\nu(w) = -\eta^{\mu\nu} \ln(z-w) \period \nn\\
\end{eqnarray}
The energy-momentum tensor is of the form 
\begin{eqnarray}
T^{EPS} &=& -\half \del x^\mu \del x_\mu -p_\al \del\theta^\al 
 -\omega_\al \del \lam^\al -b_P \del c_\Ptil \comma 
 \label{TEPS}
\end{eqnarray}
with the total central charge vanishing. Finally, the $B$-ghost field is 
 given by 
\begin{eqnarray}
B &=& -\omega_\al \del \theta^\al + \half b_P \Pi^-_\Ptil \period
\label{Bghost}
\end{eqnarray}
%%%%%%%%%%%%%%%%%%%%%%%%%%%%%%%%%%%%%%%%%
\section{Operator Mapping between BRST and Light-Cone Formalisms for 
Bosonic String }
%%%%%%%%%%%%%%%%%%%%%%%%%%%%%%%%%%%%%%%%
In order to gain  insights into our main problem of connecting 
 EPS with LCGS, 
 we study in this section a simpler problem of mapping the BRST formalism 
into the light-cone formalism 
via a similarity transformation in bosonic string. Although the 
equivalence of these two formalisms was elucidate long ago in the 
 seminal work of Kato and Ogawa \cite{KatoOgawa}, 
the explicit operator mapping constructed below, to our knowledge, is 
new\footnote{A general discussion of a similarity transformation of the 
 kind is given  by W.~Siegel in his web-published book ``Fields''.}. 
%%%%%%%%%%%%%%%%%%
\subsection{From BRST to Light-Cone}
%%%%%%%%%%%%%%%%%%
Let us recall that the BRST charge $Q$ for the bosonic string takes the form
 \cite{KatoOgawa}
\begin{eqnarray}
Q &=& \sum c_{-n} L_n -\half \sum (m-n): c_{-m}c_{-n}b_{m+n} :\comma 
\end{eqnarray}
where the symbol $:\ \ :$ stands for normal-ordering and 
the Virasoro generators  are given by 
\begin{eqnarray}
L_n &=& \half \sum_{k \in Z} \al^\mu_{n-k} \al_{\mu k} \comma \quad 
(n \ne 0) \comma \\
L_0 &=& \half p^2 + \sum_{m\ge 1} \al_{-m}^\mu \al_{\mu m} -1 \period
\end{eqnarray}
The oscillators for the non-zero modes satisfy the commutation relations
 $\com{\al^\mu_m}{\al^\nu_n} = m\eta^{\mu\nu} \delta_{m+n,0}$ with 
 the metric convention $\eta_{\mu\nu} = (-,+,+,\ldots, +)$ 
and we have set $\al' = 1/2$. 
For any Lorentz vector $A^\mu$, we define the light-cone components as
$A^\pm \equiv  {1\over \sqrt{2}} (A^0 \pm A^{25}) $
so that 
$A^\mu B_\mu = -(A^+B^- +A^-B^+) + A^i B^i \comma  i=1\sim 24 $. 
Then  $\al^\pm_n$ satisfy the commutation relations 
$\com{\al^\pm_m}{\al^\mp_n} = -m \delta_{m+n,0}\comma 
 \com{\al^\pm_m}{\al^\pm_n} =0$. 

In the light-cone formalism, the set of non-zero modes 
$(b_n, c_n, \al^+_n, \al^-_n)_{n\ne 0}$ are absent. In the BRST frame work
 it means that they must form a ``quartet'' with respect to an appropriate 
nilpotent operator, to be called  $\delta$, and decouple from 
 the cohomology of $Q$. Such a $\delta$ can easily be found in $Q$. 
When $p^+\ne 0$, $\sum c_{-n}L_n$ contains the operator 
\begin{eqnarray}
\delta &\equiv & -p^+ \sum_{n\ne 0} c_{-n} \al^-_n \comma 
\end{eqnarray}
and we have
\begin{eqnarray}
\acom{\delta}{\delta} &=& 0 \comma \\
\com{\delta}{\al^+_{-n}} &=& p^+ n c_{-n} \comma \qquad 
 \acom{\delta}{c_{-n}} =0 \comma \\
\acom{\delta}{b_{-n}} &=& - p^+\al^-_{-n} \comma \qquad 
\com{\delta}{\al^-_{-n}} = 0 \period
\end{eqnarray}
This shows that indeed $(b_n, c_n, \al^+_n, \al^-_n)_{n\ne 0}$ form a quartet 
 (\ie two doublets) 
 with respect to $\delta$. (When $p^-\ne 0$, we may define a similar 
operator interchanging $\al^+_n \leftrightarrow \al^-_n, p^+ \leftrightarrow p^-$. In the following we consider the case $p^+\ne 0$. )

 Our aim is to demonstrate this decoupling in a manifest manner by 
constructing a similarity transformation
\begin{eqnarray}
\Qbar &=& e^T Q e^{-T} \comma 
\end{eqnarray}
where $T$ is a suitable operator and
\begin{eqnarray}
\Qbar &=& \delta + Q_{lc} \comma \\
Q_{lc} &=& c_0 L^{lc}_0 \comma \label{Qlc}\\
L_0^{lc} &=& \half p^2 + \sum_{n \ge 1} \al^i_{-n} \al^i_n -1 \period
\label{Lzerolc}
\end{eqnarray}
As is almost evident from this form, 
 the cohomology 
 of $\Qbar$ consists of the usual 
light-cone on-shell states satisfying
 $L_0^{lc}\ket{\psi} =0$ (modulo inessential double degeneracy due to 
$c_0$ mode). More precise discussion will be given at the end of this 
 subsection. 

To find such a 
transformation systematically, it is convenient to distinguish 
 the members of the quartet by assigning  non-vanishing degrees
  to them in such a way that (i) $\delta$ will carry 
 degree $-1$ and (ii) the remaining part of $Q$ will carry 
 non-negative degrees. Such an assignment is given by\footnote{ 
This grading is essentially a refined version of the one used in 
\cite{KatoOgawa}.}
\begin{eqnarray}
{\rm deg}\, (\al_n^+) &=& 2\comma \qquad {\rm deg}\, (\al_n^-) =-2 \comma \\
{\rm deg}\, (c_n) &=& 1 \comma \qquad {\rm deg}\, (b_n) = -1\comma 
\qquad (n \ne 0) \\
\deg (\mbox{rest}) &=& 0 \period
\end{eqnarray}
 Then, the BRST operator $Q$ is decomposed according 
 to this degree as
\begin{eqnarray}
Q &=& \delta + Q_0 + d_1 +d_2 +d_3 \comma 
\end{eqnarray}
where the degree is indicated by the subscript, except for the 
 $\delta$ carrying  degree $-1$. 
The explicit forms of these operators are 
\begin{eqnarray}
\delta &\equiv & -p^+ \sum_{n\ne 0} c_{-n} \al^-_n \comma \label{delta}\\
Q_0 &=& c_0 L_0^{tot} = 
c_0  (L_0 + \sum_{n\ne 0} n:c_{-n} b_n: ) \comma\label{Qzero} \\
d_1 &=& \sum_{n \ne 0} c_{-n} \Ltil_n -\half \sum_{{\rm nzm}} (m-n): c_{-m}c_{-n} b_{m+n}:  \comma \\
d_2 &=& -b_0 \sum_{n \ne 0}n: c_{-n}c_n: \comma \\
d_3 &=& -p^-\sum_{n\ne 0} c_{-n} \al^+_n \comma 
\end{eqnarray}
where 
\begin{eqnarray}
\Ltil_n &\equiv & p^i \al^i_n 
+ \half \sum_{{\rm nzm}}
     \al^\mu_{n-m} \al_{\mu m} \label{defLtil}
\end{eqnarray}
and ``nzm'' indicates that only the non-zero modes are to be summed.

The advantage of the decomposition above is that the nilpotency 
 relation $\acom{Q}{Q} =0$ splits into a set of simple relations at 
 different degrees. Explicitly we have 
\begin{eqnarray}
(E_{-2}) && \acom{\delta}{\delta} =0\comma  \label{Emtwo}\\
(E_{-1}) &&  \acom{\delta}{Q_0} =0 \comma\\
(E_0) && \acom{Q_0}{Q_0} + 2 \acom{\delta}{d_1} =0 \comma\\
(E_1)&& \acom{Q_0}{d_1} +\acom{\delta}{d_2} =0\comma \label{Eone} \\
(E_2)&& \acom{d_1}{d_1} + 2\acom{Q_0}{d_2} + 2\acom{\delta}{d_3} =0\comma 
\label{Etwo}\\
(E_3)&& \acom{Q_0}{d_3} + \acom{d_1}{d_2} =0 \comma \label{Ethree}\\
(E_4)&& \acom{d_2}{d_2} + 2\acom{d_1}{d_3} =0\comma \label{Efour}\\
(E_5)&& \acom{d_2}{d_3} =0\comma \\
(E_6) && \acom{d_3}{d_3} =0\period \label{Esix}
\end{eqnarray}
Using these relations, we now show that an operator $R$ exists such that 
 $d_1, d_2$ and $d_3$ can be removed by a similarity transformation
 in the following manner:
\begin{eqnarray}
Q &=& \delta + Q_0 + d_1+d_2+d_3= e^{-R} (\delta + Q_0) e^{R} 
 \nn\\
&=& \delta + Q_0 + \com{\delta}{R} + \com{Q_0}{R}
 + \half \com{\com{\delta}{R}}{R} + \half \com{\com{Q_0}{R}}{R} +\cdots
\period \label{simtrone}
\end{eqnarray}
Since $Q_0$ is easily seen to be nilpotent, together with the 
 relations $(E_{-2})$ and $(E_{-1})$ we see that $\delta + Q_0$ is 
 nilpotent. This of course is necessary for the consistency 
 of the relation (\ref{simtrone}), but it is non-trivial to prove 
 that the series terminates at finite terms to precisely reproduce $Q$.

To find $R$ and prove (\ref{simtrone}),  knowledge of the homology 
 of the operator $\delta$ will be useful. To this end, consider the 
 operator $\Khat$ of degree 1 given by
\begin{eqnarray}
\Khat &\equiv & {1\over p^+}\sum_{n\ne 0}{1\over n}  \al^+_{-n}b_n \period
\end{eqnarray}
Further define
\begin{eqnarray}
\Nhat &\equiv & \acom{\delta}{\Khat} 
 = \sum_{n\ne 0} :( c_{-n} b_n -{1\over n} \al^+_{-n} \al^-_n ) :\period
\end{eqnarray}
It is easy to see that $\Nhat$ is an extension of the ghost number operator
 and assigns the ``$\Nhat$-number'' $(1,-1,1,-1)$ to the quartet 
$(c_n, b_n, \al^+_n, \al^-_n)$. Now let $\calO$ be a $\delta$-closed 
 operator carrying $\Nhat$-number $n$,
\ie $\gcom{\delta}{\calO} =0$ and  $\com{\Nhat}{\calO} =n \calO$. 
If $n \ne 0$, we can express $\calO$ as 
\begin{eqnarray}
\calO = (1/n) \bigl[\{\delta,\Khat\} , \calO\bigr]
= (1/n) \bigl[\delta, [\Khat,\calO\}\bigr\} \label{deltahom}
\end{eqnarray}
 and hence $\calO$ is 
$\delta$-exact. So the non-trivial homology of $\delta$ can only 
be in the sector  where $\Nhat =0$. 

%%%%%%%%%%%%%%%%%%%%%%%%%
We are now ready for the construction of $R$. Consider first the 
 relation $(E_0)$. Since $Q_0$ itself is nilpotent, it dictates
$\acom{\delta}{d_1} =0$, namely that $d_1$ is $\delta$-closed. 
But since $\com{\Nhat}{d_1} =d_1$, application of (\ref{deltahom}) 
allows us to write 
\begin{eqnarray}
d_1 &=& \com{\delta}{R_2} \comma \qquad R_2 \equiv  \acom{\Khat}{d_1} \period
\end{eqnarray}
Next, look at the relation $(E_1)$. Since $\acom{\delta}{d_2} =0$ 
holds by inspection, $(E_1)$  actually splits into two relations:
\begin{eqnarray}
\acom{\delta}{d_2} &=& 0 \comma \\
\acom{Q_0}{d_1} &=& 0 \period \label{eqE1}
\end{eqnarray}
Now since $\com{\Nhat}{d_2} =2d_2$, the former relation together with 
(\ref{deltahom}) tells us that $d_2$ can be written as
\begin{eqnarray}
d_2 &=& \com{\delta}{R_3} \comma \qquad R_3 \equiv  \half \acom{\Khat}{d_2} 
\period
\label{defR3}
\end{eqnarray}

Comparing with the expansion (\ref{simtrone}), the results 
 obtained above suggest that $R$ is given  by  $R=R_2+R_3$. Indeed the 
rest of the work is to confirm this expectation. The main points are 
\begin{itemize}
	\item $d_3$ is produced as $d_3=
\half \com{\com{\delta}{R_2} }{R_2} + \com{Q_0}{R_3} $
	\item All the unwanted commutators indeed vanish. 
\end{itemize}
Since the details are somewhat long, we relegate them to Appendix B.1. 
To summarize, we have so far shown
\begin{eqnarray}
Q &=& e^{-R} \Qtil e^R\comma \qquad \Qtil = \delta + Q_0 \comma 
\label{QQtil} \\
R &=& \acom{\Khat}{d_1+ \half d_2} \period \label{defR}
\end{eqnarray}
%%%%%%%%%%%%%%%%%%%%

Next, we shall further reduce $\Qtil$ to $\Qbar$ 
so that the light-cone structure 
 is manifestly visible. To this end, define 
an operator $\Ktil$, similar to but different from $\Khat$,  by 
\begin{eqnarray}
\Ktil &\equiv & {1\over p^+} \sum_{n\ne 0} \al^+_{-n}b_n \period
\end{eqnarray}
Then the anti-commutator with $\delta $ yields the operator
\begin{eqnarray}
\Ntil &=& \acom{\delta}{\Ktil} = 
 \sum_{n\ge 1} (-\al^-_{-n} \al^+_n - \al^+_{-n}\al^-_n + n(b_{-n}c_n
+c_{-n} b_n)) \comma \label{Ntil}
\end{eqnarray}
which evidently measures the Virasoro level of the 
 member of the quartet $q_{-n} = (\al^\pm_{-n},
 b_{-n}, c_{-n})$ in the sense
$\com{\Ntil}{q_{-n}} = n q_{-n} $. Clearly $\Ntil$ commutes with $\Qtil$. 
Further it is easy to check that, just like $\Khat$, the operator $\Ktil$ 
anti-commutes with $Q_0$. Therefore (\ref{Ntil}) gives the relation
\begin{eqnarray}
\acom{\Qtil}{\Ktil} = \Ntil \period
\end{eqnarray}
Now note that $Q_0$ given in (\ref{Qzero}) can be decomposed as
\begin{eqnarray}
Q_0 &=& c_0 L_0^{lc} + c_0 \Ntil = Q_{lc} + c_0 \Ntil \comma
\end{eqnarray}
where $L_0^{lc}$ and $Q_{lc}$ were defined in (\ref{Lzerolc})
 and (\ref{Qlc}). Therefore an additional 
 similarity transformation by $S\equiv -c_0\Ktil$ 
precisely removes the $c_0\Ntil$ part of 
 $Q_0$ and achieves our goal:
\begin{eqnarray}
\Qbar &\equiv &e^T Q e^{-T} = e^{S} e^R Q e^{-R}e^{-S}\nn\\
 &=& e^{-c_0\Ktil} \Qtil e^{c_0\Ktil}
 = \Qtil -c_0 \acom{\Ktil}{\Qtil} 
 = \Qtil -c_0\Ntil =\delta+ Q_{lc} \period \label{simQbar}
\end{eqnarray}

We now give a rather standard argument to confirm that $\Qbar$
 defines the light-cone theory. Let $\ket{\psi} $ be a $\Qbar$-closed state
 at the Virasoro level $n$ with respect to the quartet oscillators,
 \ie $\Qbar \ket{\psi} =0$ and $\Ntil \ket{\psi} =n \ket{\psi} $. 
Then, since $\acom{\Qbar}{\Ktil} =\Ntil$ holds, $n \ket{\psi} = 
\Ntil \ket{\psi} = \acom{\Qbar}{\Ktil} \ket{\psi} = \Qbar
 (\Ktil \ket{\psi} )$. Hence $\ket{\psi} $ with 
  $n\ne 0$, \ie with quartet excitations, 
 is cohomologically trivial. Disregarding such states, 
 the cohomology of $\Qbar$ is reduced to 
 that of $Q_{lc}$.  Now in the reduced space 
a general state  $\ket{\psi} $ can be written as
 $\ket{\psi} = \ket{\phi} + c_0 \ket{\chi} $, where $\ket{\phi} $ and 
$\ket{\chi} $ do not contain ghost zero modes. $Q_{lc}$-closed condition 
on $\ket{\psi} $ 
 imposes the light-cone on-shell condition $L_0^{lc} \ket{\phi} =0$ on 
 $\ket{\phi} $. On the other hand, using  $\acom{Q_{lc}}{b_0} = L_0^{lc}$
we can make a similar argument as above to conclude that if $L_0^{lc}$ 
 does not annihilate $c_0\ket{\chi} $, it can actually be written as 
a $Q_{lc}$-exact state. Thus the cohomology of $Q_{lc}$ is represented 
 by $\ket{\phi} + c_0\ket{\chi} $ with $L_0^{lc} \ket{\phi} =0$ and 
$L_0^{lc} \ket{\chi} =0$. Apparently the spectrum is doubly degenerate
 but, as is well-known, $\ket{\chi} $ does not contribute
 to the physical amplitude. Indeed the inner product of two physical 
 states $\ket{\psi} $ and $\ket{\psi'} $ is given by 
$\bra{\psi'} c_0 \ket{\psi} = \bra{\phi'} c_0 \ket{\phi} = 
\langle \phi' | \phi \rangle_{lc}$, where the subscript ``$lc$'' signifies
 the space without the ghost zero modes, and computations in 
 the light-cone theory are reproduced.

%%%%%%%%%%%%%%%%%%
\subsection{DDF operators and Virasoro Algebra}
%%%%%%%%%%%%%%%%%%%
Having succeeded in connecting
 the covariant BRST and the light-cone formalisms 
 by an explicit similarity transformation, it is natural and 
interesting to ask how various operators on both sides are mapped
 by this transformation. 

One immediate question is: What
 is the counterpart of the transverse oscillators
$\al^i_n$ on the covariant BRST side ?   Since such operators must satisfy 
 the same commutation relations as $\al^i_n$'s, it is natural to 
 guess that they must be intimately related to the DDF operators \cite{DDF}. 

Let us write the mode expansion of the open string coordinate 
$X^\mu(\sigma, \tau)$ at $\sigma=0$ as 
\begin{eqnarray}
X^\mu(\tau) &=& x^\mu + p^\mu \tau + i Y^\mu(\tau) \comma \\
Y^\mu(\tau) &=& \sum_{n\ne 0} {1 \over n} \al^\mu_n e^{-in\tau} \comma 
\end{eqnarray}
where $0 \le \tau \le 2\pi$.  
Then, the DDF operator $A^i_n$ is constructed from the photon vertex 
 operator at a special light-like momentum as 
\begin{eqnarray}
A^i_n &=& e^{inx^+/p^+} \Ach^i_n \comma \\
\Ach^i_n &\equiv &  \int [d\tau]  e^{in\tau} \Xdot^i(\tau)
 e^{-(n/p^+) Y^+(\tau)} \comma 
\end{eqnarray}
where $\int [d\tau] \equiv  \int_0^{2\pi} d\tau/2\pi$ and 
for convenience we have separated the zero-mode phase factor. 
Expanding $\Ach^i_n$ out a few terms in powers of $1/p^+$ and 
 noting that the similarity transformation by $-S=c_0\Ktil$ does not 
 affect $\al^i_n$, one is lead to 
 conjecture the simple relation
\begin{eqnarray}
\Ach^i_n &=& e^{-T} \al^i_n e^T = 
e^{-R} \al^i_n e^R =\al^i_n - \com{R}{\al^i_n} + 
\half \com{R}{\com{R}{\al^i_n} } +\cdots \period \label{DDFrel}
\end{eqnarray}

 To prove it, we need to 
know the explicit form of $R$ defined in (\ref{defR}). After a straightforward
 calculation we get
\begin{eqnarray}
R &=& R_2 + R_3 = {1\over p^+} \sum_{k\ne 0} {1\over k}\al^+_{-k} 
\Ltil^{tot}_k \comma \\
\Ltil_k^{tot} &\equiv &   \Ltil_k + \sum_{n\ne 0} nc_{-n} b_{k+n} \comma 
\end{eqnarray}
where $\Ltil_k$ was defined in (\ref{defLtil}). It is easy to show that, 
with respect to $\Ltil^{tot}_n$ operators, 
$\dot{Y}^i(\tau), \dot{Y}^+(\tau)$ and $b(\tau) = \sum_n b_n e^{-in\tau}$ 
 behave as conformal primary fields of dimension 1, while $c(\tau) 
= \sum_n c_n e^{-in\tau}$ behaves as one of dimension 0. Namely we have 
\begin{eqnarray}
\com{\Ltil^{tot}_k}{\phi(\tau)} &=& e^{ik\tau} \left( {1\over i} \del_\tau
 + k\right) \phi(\tau) \comma \\
\phi (\tau) &=& ( \dot{Y}^i(\tau), \dot{Y}^+(\tau), b(\tau))\comma  \\
\com{\Ltil_k^{tot}}{c(\tau)} &=& e^{ik\tau} {1\over i} \del_\tau
 c(\tau) \period
\end{eqnarray}
Using these properties, the action of $R$ on the basic fields is worked out as
\begin{eqnarray}
\com{R}{\dot{Y}^i(\tau)} 
&=& -{1\over p^+} {1\over i} \del_\tau (Y^+(\tau) \dot{Y}^i(\tau))\comma \\
\com{R}{Y^+(\tau)} &=& -{1\over p^+} {1\over i} Y^+(\tau) \del_\tau Y^+(\tau)
\comma  \\
\com{R}{b(\tau)} &=& -{1\over p^+} {1\over i} \del_\tau (Y^+(\tau) b(\tau))
\comma  \\
\com{R}{c(\tau)} &=& -{1\over p^+} {1\over i} Y^+(\tau) \del_\tau c(\tau)
\period
\end{eqnarray}
Now the computation of the inverse similarity transformation is 
straightforward: The single commutator is given by
\begin{eqnarray}
\com{R}{\al^i_n} &=& \int[d\tau] e^{in\tau} \com{R}{\dot{Y}^i(\tau)} 
= {n \over p^+}  \int[d\tau] e^{in\tau} Y^+(\tau) \dot{Y}^i(\tau) \period
\end{eqnarray}
Further the double commutator yields
\begin{eqnarray}
\com{R}{\com{R}{\al^i_n} } &=& {n \over p^+}  \int[d\tau] e^{in\tau}
\com{R}{ Y^+(\tau) \dot{Y}^i(\tau)}
= \left({n \over p^+} \right)^2  \int[d\tau] e^{in\tau} 
Y^+(\tau)^2 \dot{Y}^i(\tau) \period \nn\\
\end{eqnarray}
It should be clear that this  process exponentiates the 
 factor $(n/p^+) Y^+(\tau)$ and the relation (\ref{DDFrel}) is proved. 

The essential property of the DDF operators is that they commute 
 with the Virasoro and the BRST operators. This can be understood 
using our similarity transformation as follows. 
Since $A^i_n$ does not contain $(b,c)$ ghosts, we only need to show 
$\com{Q}{A^i_n} =0$. By the similarity transformation, this is 
 mapped to 
\begin{eqnarray}
\com{\delta +Q_{lc} }{e^{inx^+/p^+} \al^i_n} \period
\end{eqnarray}
 $\delta$ trivially commutes with 
$e^{inx^+/p^+} \al^i_n$. $Q_{lc}$ also commutes with it since
\begin{eqnarray}
\com{L_0^{lc}}{e^{inx^+/p^+} \al^i_n} &=& \com{L_0^{lc}}{e^{inx^+/p^+}} \al^i_n
+e^{inx^+/p^+} \com{L_0^{lc}}{ \al^i_n} \nn\\
&=& n e^{inx^+/p^+} \al^i_n -n e^{inx^+/p^+} \al^i_n =0 \period
\end{eqnarray}
Note that the phase factor in front is crucial for the cancellation. 

A related question of interest is:\ 
What happen to the Virasoro generators when transformed to 
the light-cone side? 
Consider the total Virasoro operators $L_n^{tot}$ including the ghost
 part. It is convenient to represent it as $L_n^{tot} =\acom{Q}{b_n} $
since we already know how $Q$ is similarity-transformed. 
Therefore the Virasoro operators on the light-cone side can be computed as
\begin{eqnarray}
\bar{L}_n^{tot} &=& e^S e^R L_n^{tot} e^{-R} e^{-S} 
  = \acom{\Qbar}{\bar{b}_n} \comma \\
\bar{b}_n &=& e^S e^R b_n e^{-R} e^{-S} \period
\end{eqnarray}
Transformation 
 of $b_n$ by $e^R$ 
can be performed in a manner similar to that for $\al^i_n$, with 
 the result
\begin{eqnarray}
\tilde{b}_n &\equiv & e^R b_n e^{-R} =  \int[d\tau] e^{in\tau} 
e^{(n/p^+)Y^+(\tau)}  b(\tau) \period
\end{eqnarray}
Further transformation by $e^S$ acts only on the zero mode part of $b(\tau)$
 in the integrand and gives 
\begin{eqnarray}
\bar{b}_n &=&   \int[d\tau] e^{in\tau} 
e^{(n/p^+)Y^+(\tau)} ( b(\tau) + \Ktil) \period
\end{eqnarray}
Then by a straightforward calculation we obtain
\begin{eqnarray}
\bar{L}_n^{tot} &=&
 \int[d\tau] e^{in\tau} e^{(n/p^+) Y^+(\tau)} 
\left(-n\check{c}(\tau)(b(\tau) + \Ktil) -ip^+ \dot{Y}^-(\tau)  
+ \Ntil + L_0^{lc} \right) \comma 
\end{eqnarray}
where $\check{c}(\tau) \equiv c(\tau) -c_0$. Note that
this is of the form
 $B_n + C_n L_0^{lc}$, where $B_n$ and $C_n$ consist entirely of unphysical 
 fields. Since $L_0^{lc}$ annihilates the physical states, Virasoro operators
are essentially inert in the light-cone formalism. One may say that 
 the gauge transformations are already ``used up'' in the light-cone 
 formulation or they only act on the unphysical sector. 
%%%%%%%%%%%%%%%%%%
\section{Relating EPS and Light-Cone GS Formalisms by Similarity 
Transformation }
%%%%%%%%%%%%%%%%%%%
We now make use of the experience gained in the previous section to 
construct the similarity transformation connecting the light-cone GS and 
 the EPS formalisms for superstring.
 Since the construction is  more involved 
 compared to the bosonic string case, we shall present it 
in two steps so that our strategy will be transparent. 
Conventions, to be used below, 
 for $\Ga^\mu$ matrices, spinors and their $U(5)$ and $SO(8)$ parametrizations
are explained in Appendix A. 
%%%%%%%%%%%%%%%%%%%
\subsection{Similarity transformation I}
%%%%%%%%%%%%%%%%%%%
Just as in the bosonic case, the final BRST operator on the LCGS side 
 should be of the form 
\begin{eqnarray}
\Qbar &=& \delta_{lc} + Q_{lc} \comma \label{superQbar}\\
Q_{lc} &=& c_0L_0^{lc} =c_0(L_{0,B}^{lc} + L_{0,F}^{lc})\period
\end{eqnarray}
Here $L_{0,B}^{lc}$ is the bosonic part of the on-shell operator identical to 
(\ref{Lzerolc}), while $L_{0,F}^{lc}$ is the fermionic part given by 
\begin{eqnarray}
L_{0,F}^{lc} &=& \sum_{n\ge 1} n S^a_{-n} S^a_n \comma 
\end{eqnarray}
where $S^a_n\ (a=1\sim 8)$ are the $SO(8)$-chiral spinor 
oscillators expressing the 
 physical fermionic excitations of the light-cone GS string. They
 should  satisfy the 
self-conjugate anti-commutation relations
\begin{eqnarray}
\acom{S^a_m}{S^b_{n}} &=& \delta^{ab} \delta_{m,-n} \period
\end{eqnarray}
$\delta_{lc}$ in (\ref{superQbar}) is the operator consisting of 
 ``unphysical'' oscillators, which allows us to identify and 
drop the unwanted states  from the cohomology. 
Thus our first task is to identify $S^a_n$ and construct 
appropriate $\delta_{lc}$ 
in terms of the EPS variables.

 The basic fermionic spinor fields of EPS are 
the conjugate pairs $(\theta^\al, p_\al)$, which can be decomposed 
 into $SO(8)$ chiral and anti-chiral pairs as 
$(\theta^a, p_a), (\theta^\adot, p_\adot)$. From the $SO(8)$-chiral 
 pair, one can construct two self-conjugate fields as 
\begin{eqnarray}
S^a &\equiv & {1\over \sqrt{2p^+}} (p^a + p^+ \theta^a) \comma \qquad 
\Stil^a \equiv  {1\over \sqrt{2p^+}} (p^a - p^+ \theta^a) \period
\label{defS}
\end{eqnarray}
Their modes satisfy the anti-commutation relations
\begin{eqnarray}
\acom{S^a_m}{S^b_{n}} &=& \delta^{ab} \delta_{m,-n} 
\comma \qquad \acom{\Stil^a_m}{\Stil^b_{n}} = -\delta^{ab} \delta_{m,-n}
\period
\end{eqnarray}
Thus, it is natural to identify $S^a$ in (\ref{defS}) as the physical 
 fermionic field of the GS formalism while the
 negatively-normed $\Stil^a$ operator, together with the anti-chiral 
 pair  $(\theta^\adot, p_\adot)$, should be regarded as unphysical. 

There are also five pairs of fermionic ghosts $(b_P, c_\Ptil)_{P =1\sim 5}$
 in the EPS formalism, which will be divided into $(b_p,c_\ptil)_{p=1\sim 4}$
 and $(b_5, c_\fivetil)$. Since the light-cone fields $\del x^\pm$ 
 should be removed together with a ``$(b,c)$'' ghost pair  just as in the 
bosonic string, it is natural to make the identification 
 $(b,c) = (\xi b_5, \xi^{-1} c_\fivetil)$, where the rescaling factor $\xi$
 will be appropriately chosen later. 

We can now depict the correspondence between the degrees of freedom 
 of  light-cone GS and  EPS by 
 the following diagram:
\begin{eqnarray}
&&\qquad \qquad \ \del x^i \oplus \ub{\del x^\pm \oplus
 (b, c)}_{\rm quartet} \comma \\
&& \nn\\
&& \overbrace{S^a \oplus \Stil^a 
\oplus (\theta^\adot, p_\adot)}^{(\theta^\al, p_\al)} \oplus
(b_p, c_\ptil)\oplus (\lam^\al, \omega_\al) \period
\end{eqnarray}
\vspace{-1.8cm}
\begin{eqnarray}
 \qquad \phantom{S^a \oplus\ \ \ } \ub{\phantom{\Stil^a 
\oplus (\theta^\adot, p_\adot) \oplus
(b_p, c_\ptil)\oplus (\lam^\al, \omega_\al)}}_{\rm quartets} \nn
\end{eqnarray}
The structure in the upper line is precisely that of the bosonic string. 
In the lower line, one sees that, as far as the counting of the 
degrees of freedom is concerned, the bosonic fields $(\lam^\al, \omega_\al)$
 and the fermionic fields $(\Stil^a, \theta^\adot, p_\adot, b_p, c_\ptil)$ 
can form 16 unphysical quartets so that only $S^a$ 
 will be left as physical, as we wish.

How the quartets are formed is dictated by the structure of the operator 
$\delta_{lc}$. It is clear that to remove the quartet 
$(\del x^\pm, b, c)$ we may use the same structure as 
 in the bosonic case, namely $\delta_{b} \equiv 
-p^+\sum_{n\ne 0} c_{-n}
\al^-_n$. It should also be evident that the   
sets $(\lam^\adot, \omega_\adot, \theta^\adot, p_\adot)$ consisting 
 of $SO(8)$ conjugate spinors form 
eight quartets with respect to the nilpotent operator $
\delta_{c}\equiv\sum_n \lam^\adot_{-n} p_{\adot,n}$. Indeed, we have 
\begin{eqnarray}
\acom{\deltac}{\theta^\adot_n} &=& \lam^\adot_n \comma \qquad 
\com{\deltac}{\lam^\adot_n} = 0\comma  \\
\com{\deltac}{\omega_{\adot,n}} &=& p_{\adot, n}\comma \qquad 
\acom{\deltac}{p_{\adot,n}} =0 \period
\end{eqnarray}

We are still left with the remaining unphysical fields
 $(\Stil^a, b_p, c_\ptil, \lam^a, \omega_a)$ to deal with.  The most 
 natural way to form quartets is to first construct, 
 out of $\Stil^a$ and $(b_p, c_\ptil)$, eight conjugate pairs 
$(\Stil^{+a},\Stil^{-a})$ satisfying the anti-commutation relations 
\begin{eqnarray}
\acom{\Stil^{+a}_m}{\Stil^{+b}_n} &=& \acom{\Stil^{-a}_m}{\Stil^{-b}_n} =0 
\comma \qquad \acom{\Stil^{+a}_m}{\Stil^{-b}_n} = \delta^{ab} \delta_{m+n,0}
\period \label{acomStilpm}
\end{eqnarray}
Then, the fields $(\Stil^{+a},\Stil^{-a}, \lam^a, \omega_a)$ form 
eight quartets with respect to the operator 
$\delta_s \equiv -2 \sum_n \lam^a_{-n} \Stil^{-a}_n$. (The factor 
 of $-2$ in front is for later convenience.) 

 To construct
$(\Stil^{+a},\Stil^{-a})$, we first reshuffle $(b_p, c_\ptil)$ to 
 form self-conjugate fields $g_a$ in the following way. Let $u_\adot$ 
 be a constant conjugate spinor normalized as $u_\adot u_\adot =2$ and form
\begin{eqnarray}
g_a &\equiv & \sqrt{p^+}\, e^{+i}_p \ga^i_{a\adot} u_\adot c_\ptil +
{1\over \sqrt{p^+}}  e^{-i}_\ptil
\ga^i_{a\adot}  u_\adot b_p \period \label{defga}
\end{eqnarray}
Here $\ga^i_{a\adot}$ are the $SO(8)$ Pauli matrices and 
$(e^{+ i}_p, e^{-i}_\ptil)$ are the $U(4)$ projectors defined 
just like the $U(5)$ counterparts given in (\ref{defeproj}). Then from the 
 anti-commutation relation $\acom{b_p}{c_\qtil} =\delta_{pq}$,  one 
can  check that $g_a$ is  self-conjugate, \ie 
 $\acom{g_a}{g_b} = \delta_{ab}$. The definition (\ref{defga}) can be 
 inverted to yield
\begin{eqnarray}
b_p &=& \sqrt{p^+}\, e^{+i}_p (\ga^i u)_a g_a \comma \qquad 
c_\ptil = {1\over \sqrt{p^+}} e^{-i}_\ptil (\ga^i u)_a g_a \period
\end{eqnarray}
Furthermore, one can easily prove the relation
\begin{eqnarray}
g_{a, -n} g_{a,n} &=& b_{p,-n} c_{\ptil,n} + c_{\ptil,-n} b_{p,n} \comma 
\label{gaga}
\end{eqnarray}
which holds regardless of the explicit choice of $u_\adot$. Using $g_a$ 
 we now define $(\Stil^{+a}_n,\Stil^{-a}_n)$ as
\begin{eqnarray}
\Stil^{+a}_n &\equiv & {1\over \sqrt{2p^+}} (\Stil^a_n + g^a_n) \comma \qquad 
 \Stil^{-a}_n \equiv \sqrt{{p^+ \over 2}} (-\Stil^a_n + g^a_n) \period
\end{eqnarray}
Then these oscillators correctly satisfy the anti-commutation relations 
 (\ref{acomStilpm}). 

Summarizing, the appropriate $\delta_{lc}$ is given by 
\begin{eqnarray}
\delta_{lc} &=& \delta_b + \delta_c + \delta_s \comma \\
\delta_b &=& -p^+\sum_{n\ne 0} c_{-n} \al^-_n 
=  -i\int[dz] p^+ (c(z) \del x^-(z))_{\rm nzm}
\comma \\
 \delta_c &=&\sum_n \lam^\adot_{-n} p_{\adot,n} =\int[dz] \lam^\adot(z)
p_\adot(z)
\comma \\
 \delta_s &=&-2 \sum_n \lam^a_{-n} \Stil^{-a}_n 
= -2 \int[dz] \lam^a(z) \Stil^{-a}(z)
\period
\end{eqnarray}
where the subscript ``nzm'' means omitting the zero mode part. 
Just as in the bosonic string case, we can easily construct 
 the homotopy operator $K$ with respect to $\delta_{lc}$ such that 
 $\acom{\delta_{lc}}{K} =N $, where $N$  is the level-counting operator
 for the unphysical modes. It is given by
\begin{eqnarray}
K &=& {1\over p^+} \sum_{n\ne 0} \al^+_{-n} b_n 
 + \sum_{n \ne 0} n (\theta^\adot_n \omega_{\adot, -n} 
 -\half \Stil^{+a}_n \omega_{a,-n} ) \period
\end{eqnarray}
We may now follow the process employed in (\ref{simQbar}) backwards and 
 add the operator $N$ to $\Qbar$,  to get a ``more covariantized'' form. 
Using the definition of $\Stil^{\pm a}$ as well as  the relation (\ref{gaga}), 
we get 
\begin{eqnarray}
\Qbar' &\equiv & e^{c_0 K} \Qbar e^{-c_0K} = \Qbar + N = \delta_{lc} + Q_0 \comma \\
Q_0 &=& c_0 L_0^{tot} \comma \\
L_0^{tot} &=& \half p^\mu p_\mu + \sum_{n\ge 1} \al^\mu_{-n}\al_{\mu n} 
 + \sum_{n\ge 1} n \biggl( p_{\al,-n} \theta^\al_n 
 + \theta^\al_{-n} p_{\al,n} \nn\\
&& + \lam^\al_{-n} \omega_{\al,n}
-\omega_{\al,-n} \lam^\al_n 
 + c_{\Ptil,-n} b_{P,n} + b_{P,-n} c_{\Ptil,n} \biggr) \period
\end{eqnarray}
It is important to note that $L_0^{tot}$ above is precisely 
 the ``zero mode part'' of the energy-momentum tensor $T^{EPS}$ 
of the EPS formalism given in (\ref{TEPS}). Thus, from now on we shall 
 denote it by $T^{EPS}_0$. 

Before we describe the main part of the 
 similarity transformation, let us make one further manipulation
 to bring $\Qbar'$ to a yet better form. Let us write out $\delta_s$ 
more explicitly. It reads
\begin{eqnarray}
\delta_s &=& -2 \int[dz] \lam^a(z) \Stil^{-a}(z) 
= \int[dz] \left(\lam^a p_a -p^+\lam^a\theta_a -\sqrt{2p^+}\,\lam^a g_a \right)
\period
\end{eqnarray}
A non-trivial fact is that this is produced from $\int[dz] \lam^a p_a$ 
by the following simple similarity transformation, as can be readily verified:
\begin{eqnarray}
\delta_s &=& e^T \int[dz] \lam^a p_a e^{-T} \comma \\
T &=& \sqrt{2p^+}\, \int [dz] \theta^a g_a \period
\end{eqnarray}
Since $T$ commutes with the rest of $\Qbar'$, we can now combine 
$\lam^\adot p_\adot$ in $\delta_c$ and $\lam^a p_a$ above to produce the 
covariant form $\lam^\al p_\al$. In this way we get
\begin{eqnarray}
\Qtil &\equiv & e^{-T} \Qbar' e^T = \int[dz]\left( \lam^\al p_\al 
-ip^+ (c \del x^-)_{nzm}\right)
 + c_0 T^{EPS}_0 \period \label{Qtil}
\end{eqnarray}
Notice that the first term $\lam^\al p_\al$ is precisely the combination 
 that appears in $\lam^\al d_\al$, the main part of the BRST-like 
 operator $\Qhat$ for EPS. Thus the remaining task is to construct 
 the rest of the similarity transformation which converts $\Qhat$ to $\Qtil$. 
%%%%%%%%%%%%%%%%%%%
\subsection{Similarity transformation II}
%%%%%%%%%%%%%%%%%%%
Recall the form of $\Qhat$:
\begin{eqnarray}
\Qhat &=& \delta + Q + d_1 +d_2 \nn\\
&=& -ib_P\Phi_\Ptil + \lam^\al d_\al + c_\Ptil \calP_P -{i \over 2}
c_\Ptil c_\Qtil \calR_{PQ} \period
\end{eqnarray}
Here and hereafter, we shall omit the integral symbol $\int[dz]$. 
Comparing it to $\Qtil$ in (\ref{Qtil}), one observes the following:
\begin{itemize}
	\item The first term $\delta =-ib_P\Phi_\Ptil$ and the last term $
d_2=-{i \over 2}
c_\Ptil c_\Qtil \calR_{PQ}$ must be removed completely. 
	\item $c_\fivetil \calP_5$ part of $d_1=c_\Ptil \calP_P$ contains the 
 structure $\sim (c \del x^-)_{\rm nzm}$ appearing in $\Qtil$. 
We must remove the 
rest, including the zero mode part $(c \del x^-)_0$. 
	\item The structure $c_0 T^{EPS}_0$ must be generated.
	\item As for $Q=\lam^\al d_\al$, we must keep the $\lam^\al p_\al$ part and 
 remove the rest. 
\end{itemize}
Below we will perform these operations  by a series of similarity
 transformations.  

The first step is to remove $\delta$ by a similarity transformation 
 of the form $e^X \Qhat e^{-X}$. We can achieve this by an operator $X$ 
 with the property $XQ =-\delta$. It is given by 
\begin{eqnarray}
X &=& -i b_P X_\Ptil\comma \qquad X_\Ptil = \calNbar^\mu_\Ptil \theta \ga_\mu
\lam\period
\end{eqnarray}
Apart form shifting $Q$ by $-\delta$, this similarity transformation 
 translates $c_\Ptil$ by $-iX_\Ptil$, as is clear from the structure 
 of $X$. In this way we get
\begin{eqnarray}
Q^{[1]} & \equiv &e^X \Qhat e^{-X} = Q + (c_\Ptil -iX_\Ptil) \calP_P
 -{i \over 2}
(c_\Ptil -iX_\Ptil)(c_\Qtil -iX_\Qtil)\calR_{PQ} \period
\end{eqnarray}

As the second step, we remove $c_\ptil \calP_p$. This can be done by 
 a similarity transformation $e^YQ^{[1]} e^{-Y}$ with 
\begin{eqnarray}
Y &=& -2i \lampinv c_\ptil d_p \comma \label{defY}
\end{eqnarray}
which, apart from a factor of 2, is very similar to the transformation
 used in \cite{AK2} 
 to connect $\Qhat$ with the Berkovits' BRST-like operator $Q_B$.
 Although 
 rather tedious, the calculation 
is straightforward. It turns out that
the net result is  simply to remove $c_\ptil$ from $Q^{[1]}$. Thus we 
 obtain
\begin{eqnarray}
Q^{[2]} &\equiv &e^Y Q^{[1]} e^{-Y} =
 Q + c_\fivetil (\calP_5 -X_\ptil \calR_{5p})
-iX_\Ptil \calP_P + {i \over 2} X_\Ptil X_\Qtil \calR_{PQ} \period
\end{eqnarray}

At this stage,
 let us try to generate the $c_0T^{EPS}_0$ structure
 appearing in (\ref{Qtil}). This step was 
 not needed in the bosonic string case, since the BRST operator already 
 contained such a term. On the other hand, in the EPS (and PS)
 formalism, Virasoro generators are not present in $\Qhat$ and they can 
 only appear through the fundamental relation $\acom{\Qhat}{B(z)} 
=T^{EPS}(z)$. 
To construct the similarity transformation that produces $c_0T^{EPS}_0$, 
we need to specify the precise definition of $(b,c)$ ghosts in terms 
 of $(b_5, c_\fivetil)$. As already mentioned, $c_\fivetil \calP_5$ term 
 contains the structure $\sim c\del x^-$ in the 
form $-4c_\fivetil \del x^+_5$.
For this to be equal to $-ip^+c \del x^-$, 
with our convention $\del x^+_5 = e^{+\mu}_5 \del x_\mu =-\del x^-/\sqrt{2}$, 
the identification should be
\begin{eqnarray}
c_\fivetil &=& -{i \over 2\sqrt{2}} p^+c \comma \qquad 
 b_5 = {i2\sqrt{2} \over p^+}b\period \label{bcfive}
\end{eqnarray}

A characteristic property of the energy-momentum tensor is that 
it is unchanged under any similarity transformation $e^U T e^{-U}$
 as long as the 
total Virasoro level of the exponent $U$ is zero. Since all the 
similarity transformations we have performed are of this type, we have
$\acom{Q^{[2]}}{B^{[2]}(z)} =T^{EPS}(z)$, where $B^{[2]}(z)$
 is the similarity-transform 
 of the B-ghost operator at this stage. Then, using its zero mode
 $B^{[2]}_0=\int[dz]z^2 B^{[2]}(z)$, a similarity transformation 
 with the exponent $c_0 B^{[2]}_0$ would produce $c_0 T_0^{EPS}$. 
This, however, is not quite the correct procedure. As  $B^{[2]}_0$ contains 
$b_0$ mode, the exponent has a term  $\sim c_0b_0$  and 
 the similarity transformation produces an infinite series. 
Thus, we must first subtract off the $b_0$($\propto b_{5,0}$)-dependent 
part from $B^{[2]}_0$. 
The original $B$ field given in (\ref{Bghost}) contains the 
$b_5$-dependent part 
\begin{eqnarray}
{1\over 2\sqrt{2}} b_5 (\del x^+ -i\sqrt{2} \theta^a \del \theta^a)
\period \label{bfiveinB}
\end{eqnarray}
 The first similarity transformation by $e^T$ does not modify this structure.
On the other hand, the second transformation by $e^X$ adds the 
contribution 
\begin{eqnarray}
X (-\omega_a \del \theta^a) = {i \over 2} b_5 \theta^a \del \theta^a
\comma 
\end{eqnarray}
which is seen to precisely cancel the second term of (\ref{bfiveinB}). 
Since the subsequent transformation by $e^Y$ does not produce any new term 
 containing $b_5$, we find that the $b_0$-dependent part of $B^{[2]}_0$ 
 is given   by $(1/2\sqrt{2})b_{5,0} (\del x^+)_0$, which  
according to the identification (\ref{bcfive}) is exactly equal 
 to $b_0$.  Therefore, 
 the appropriate transformation is of the form $e^Z$ with 
\begin{eqnarray}
Z &=& c_0 \check{B}_0 \comma \qquad \check{B}_0 \equiv B^{[2]}_0 -b_0
\period
\end{eqnarray}
A simple calculation then yields 
\begin{eqnarray}
Q^{[3]} &\equiv & e^Z Q^{[2]} e^{-Z} = Q^{[2]} + c_0 T^{EPS}_0 
 -c_{5,0} (\calP_5 - X_\ptil \calR_{5p})_0 \period
\end{eqnarray}
Note that apart from the desired term $c_0T^{EPS}_0$, 
the above transformation also generated the last term, which
 comes from $-c_0 \acom{b_0}{Q^{[2]}} = -c_{\fivetil,0} 
\acom{b_{5,0}}{Q^{[2]}} $. 
It subtracts off 
 from the term $c_5  (\calP_5 - X_\ptil \calR_{5p})$ in $Q^{[2]}$ 
the part proportional to $c_{5,0}$. This turns out to be quite significant. 
As already pointed out at the beginning of this subsection, 
the zero mode part of the term $-4c_\fivetil \del x_5^+=-ip^+c
\del x^-$ in $c_\fivetil \calP_5$  must be removed in order to 
 obtain the correct structure we need in $\Qtil$ of (\ref{Qtil}). What we 
 have found is that the subtraction of 
$b_0$ from $B^{[2]}_0$ needed to correctly generate 
the  $c_0T^{EPS}_0$ structure simultaneously does this job for us. This is 
 a subtle but compelling indication that our similarity transformation 
is on the right track. 

What remains to be done is to remove the rest of the unwanted
 fields by judicious similarity transformations 
and bring $Q^{[3]}$ into $\Qtil$ given in (\ref{Qtil}). 
We will  achieve this in two steps. 
 
Consider first the decoupling of 
 $(\lambda_{p5},\omega_{\ptil\tilde{5}},\theta_{p5},p_{\ptil\tilde{5}})$, 
 which form a quartet with respect to the operator $\delta^{[3]} 
 \equiv \lambda_{p5}p_{\ptil\tilde{5}}$ 
 contained in $Q^{[3]}$. For this purpose, 
 we assign the following degrees to the members of the quartet:
\begin{align}
\deg(p_{\ptil\tilde{5}}) &= -2\comma \quad \deg(\theta_{p5}) = +2\comma \\
\deg(\omega_{\ptil\tilde{5}}) &= -1\comma \quad \deg(\lambda_{p5}) 
= +1\comma \\
\deg(\text{rest}) &= 0 \period
\end{align}
Under this grading, $\delta^{[3]}$ has degree $-1$ and 
$Q^{[3]}$ splits into 
\begin{align}
Q^{[3]} &= \delta^{[3]} + Q^{[3]} _{0}
  + d^{[3]}_{1} + d^{[3]}_{2} + d^{[3]}_{3} + d^{[3]}_{4} \comma
\end{align}
where $(Q^{[3]}_{0},d^{[3]}_{n})$ carry
 degrees $(0,n)$ respectively. What we wish to do is to retain the 
 first two terms 
\begin{align}
\delta^{[3]} &= \lambda_{p5}p_{\ptil\tilde{5}} \comma \\
Q^{[3]}_{0} 
 &= [\lambda^{\alpha}p_{\alpha} - \lambda_{p5}p_{\ptil\tilde{5}}]
  - 4(c_{\tilde{5}}\delx^{+}_{5})_{\text{nzm}} \nonumber\\
 &\quad -\lambda_{pq}(\theta_{+}\theta_{\ptil}\del\theta_{\qtil}
       + 2i\lambda_{+}^{-1}\theta_{+}\lambda_{\ptil}\delx^{-}_{\qtil}) \nonumber\\
 &\quad -\theta_{pq}(\theta_{+}\lambda_{\ptil}\del\theta_{\qtil}
       -\theta_{+}\theta_{\ptil}\del\lambda_{\qtil}       
       +\lambda_{+}\theta_{\ptil}\del\theta_{\qtil}
       -2i\lambda_{\ptil}\delx^{-}_{\qtil}) \nonumber\\
 &\quad + c_{0}T^{EPS}_{0} \comma
 \label{eq:defQ3-0}
\end{align}
and remove the rest carrying degree $\ge 1$. 
Although it is straightforward to write down the explicit forms 
 of $d^{[3]}_n$, what will be important is only their basic structure. 
Making explicit the dependence on $\lam_{p5}$ 
and $\theta_{p5}$, we have
\begin{align}
d^{[3]}_{1} &= \lambda_{p5}F_{\ptil} \comma
 \label{eq:structd3-1}\\
d^{[3]}_{2} &= \theta_{p5}B_{\ptil} 
  + \lambda_{p5}\lambda_{q5}F'_{\ptil\qtil} \comma
 \label{eq:structd3-2} \\
d^{[3]}_{3} &= \lambda_{p5}\theta_{q5}B_{\ptil\qtil} 
  + \lambda_{p5}\del\theta_{q5}\tilde{B}_{\ptil\qtil} \comma
 \label{eq:structd3-3} \\
d^{[3]}_{4} &= \theta_{p5}\theta_{q5}F_{\ptil\qtil}
  + \theta_{p5}\del\theta_{q5}\tilde{F}_{\ptil\qtil} \comma
 \label{eq:structd3-4}
\end{align}  
where $F$'s and  $B$'s are, respectively, fermionic and bosonic 
 expressions free of the quartet members. Due to this property, 
they (anti-)commute with $\delta^{[3]}$ and with themselves. 
This fact will be utilized extensively below. 

To remove $d^{[3]}_1 \sim d^{[3]}_4$, we need to find an operator 
 $R$ such that 
\begin{align}
 Q^{[3]} &= \delta^{[3]} + Q^{[3]}_{0} 
   + d^{[3]}_{1} + d^{[3]}_{2} + d^{[3]}_{3} + d^{[3]}_{4} \nonumber\\
 &= e^{-R} (\delta^{[3]}+Q^{[3]}_{0}) e^R
 \label{eq:simQ3toQ4}
\end{align}
is realized. It is easy to see that $R$ must start with degree two 
 and hence its degree-wise decomposition can be written as 
$R=R_2+R_3+ \cdots$. 
 Then, the decomposition of 
 the equation (\ref{eq:simQ3toQ4}) with respect to the degree 
 is given by
\begin{align}
d^{[3]}_{1} 
 &= \bigl[\delta^{[3]},R_{2}\bigr] \comma
 \label{eq:simQ3toQ4deg1}\\
d^{[3]}_{2} 
 &= \bigl[\delta^{[3]},R_{3}\bigr] 
   +\bigl[Q^{[3]}_{0},R_{2}\bigr] \comma
 \label{eq:simQ3toQ4deg2}\\
d^{[3]}_{3} 
 &= \bigl[\delta^{[3]},{R}_{4}\bigr] 
  + \bigl[Q^{[3]}_{0},R_{3}\bigr]
  + {1\over2}\bigl[\bigl[\delta^{[3]},R_{2}\bigr],R_{2}\bigr] \comma
 \label{eq:simQ3toQ4deg3}\\
d^{[3]}_{4} 
 &= \bigl[\delta^{[3]},R_{5}\bigr] 
  + \bigl[Q^{[3]}_{0},R_{4}\bigr] \nonumber\\
&\qquad
  + {1\over2}\bigl[\bigl[\delta^{[3]},R_{2}\bigr],R_{3}\bigr]
  + {1\over2}\bigl[\bigl[\delta^{[3]},R_{3}\bigr]R_{2}\bigr]
  + {1\over2}\bigl[\bigl[Q^{[3]}_{0},R_{2}\bigr],R_{2}\bigr] \comma
 \label{eq:simQ3toQ4deg4}\\
0 &= \bigl[\delta^{[3]},R_{6}\bigr] + \cdots \comma
 \label{eq:simQ3toQ4deg5}
\end{align}
and so on. Just as we did for the bosonic string, together with the 
relations that follow from the nilpotency of $Q^{[3]}$ 
 we can solve these
 equations successively and determine $R$. 

Let us illustrate the first step. To solve the equation 
(\ref{eq:simQ3toQ4deg1}) at degree 1, we look at the degree 0 part of 
 $\acom{Q^{[3]}}{Q^{[3]}} =0$, which 
 reads  $\{Q^{[3]}_{0},Q^{[3]}_{0}\} + 2\{\delta^{[3]},d^{[3]}_{1}\}=0$.
 Since $Q_0^{[3]}$ itself is nilpotent, 
we must have  $\{\delta^{[3]},d^{[3]}_{1}\}=0$. 
This suggests that $d^{[3]}_{1}$ may be written as 
 $d^{[3]}_{1} = [\delta^{[3]},R_{2}]$, with a suitable $R_2$. 
 From the structure of $\delta^{[3]}$
 and $d^{[3]}_{1}$ it is easy to see that 
$ R_{2} = \theta_{p5}F_{\ptil} $
is the desired operator. 

The procedure to solve the equations at higher degrees is similar:
 The nilpotency of $Q^{[3]}$ at degree $n-1$ suggests
 the existence of a solution to the equation (\ref{eq:simQ3toQ4}) 
 at degree $n$,  and  $R_{n}$ can then be obtained
 using the specific  structure of $d^{[3]}_{n}$'s
 given in (\ref{eq:structd3-1})$\sim$ (\ref{eq:structd3-4}). 
Although this process is not in general guaranteed 
 to terminate at finite steps, it does so in the present case, 
with the result 
\begin{eqnarray}
R &=& R_2+R_3+R_4 \comma \\
R_2 &=& \theta_{p5}F_{\ptil} \comma \label{defR2}\\
R_{3} &=& \lambda_{p5}\theta_{q5}F'_{\ptil\qtil} \comma \\
R_{4} &=& {1\over2}\theta_{p5}\theta_{q5}B^{A}_{\ptil\qtil}
 + {1\over2}\theta_{p5}\del\theta_{q5}\tilde{B}^{S}_{\ptil\qtil}
\comma 
\end{eqnarray}
where the superscripts $A$ and $S$ signify the antisymmetric and the 
symmetric parts. Some details leading to this result are given
 in Appendix B.2.

Having decoupled $\lam_{p5}$ and $\theta_{p5}$, the next step is 
 to decouple  $(\lambda_{pq},\omega_{\ptil\qtil},\theta_{pq},p_{\ptil\qtil})$.
This set of fields form a  quartet with respect to the operator 
$\delta^{[4]} \equiv {1\over2}\lambda_{pq}p_{\ptil\qtil}$,
 contained in $Q^{[4]} \equiv \delta^{[3]}+Q^{[3]}_{0}$, which is
 the result of the previous step. The strategy should now be familiar. 
By assigning the degrees 
\begin{align}
\deg(\lambda_{pq}) &= 1 \comma \quad \deg(\omega_{\ptil\qtil}) = -1 \comma \\
\deg(\theta_{pq}) &= 2 \comma \quad \deg(p_{\ptil\qtil}) = -2 \comma\\
\deg(\text{rest}) &= 0 \comma
\end{align}
we can decompose $Q^{[4]}$ into 
\begin{align}
Q^{[4]} &= \delta^{[4]} + Q^{[4]}_{0} + d^{[4]}_{1} + d^{[4]}_{2} \comma 
\end{align}
where $\deg\,(\delta^{[4]},Q^{[4]},d^{[4]}_{n}) = (-1,0,n)$. 
The explicit forms of these structures are 
\begin{align}
\delta^{[4]} 
 &= {1\over2}\lambda_{pq}p_{\ptil\qtil} \comma \\
Q^{[4]}_{0} 
 &= [\lambda^{\alpha}p_{\alpha} - {1\over2}\lambda_{pq}p_{\ptil\qtil}]
 - 4(c_{\tilde{5}}\delx^{-}_{\tilde{5}})_{\text{nzm}}
 + c_{0}T^{EPS}_{0} \comma \\
d^{[4]}_{1} 
 &= -\lambda_{pq}(\theta_{+}\theta_{\ptil}\del\theta_{\qtil}
       + 2i\lambda_{+}^{-1}\theta_{+}\lambda_{\ptil}\delx^{-}_{\qtil})
 \comma \\
d^{[4]}_{2}
 &= -\theta_{pq}(\theta_{+}\lambda_{\ptil}\del\theta_{\qtil}
       -\theta_{+}\theta_{\ptil}\del\lambda_{\qtil}       
       +\lambda_{+}\theta_{\ptil}\del\theta_{\qtil}
       -2i\lambda_{\ptil}\delx^{-}_{\qtil}) \period
\end{align}
At this stage, it is gratifying to note that the sum 
$\delta^{[4]} + Q^{[4]}_{0}$ 
is precisely the expression $\Qtil$ that we want. 
Thus, the only remaining task is to show that 
 $d^{[4]}_{1}$ and $d^{[4]}_{2}$ 
 can be removed by a similarity transformation of the form 
\begin{align}
 Q^{[4]} &= \delta^{[4]} + Q^{[4]}_{0} + d^{[4]}_{1} + d^{[4]}_{2} 
=  e^{-S} (\delta^{[4]} + Q^{[4]}_{0} )   e^{S}  = e^{-S} \Qtil e^S \period
\label{Stransf}
\end{align}
by a suitable operator $S$. A straightforward computation shows that 
 $d^{[4]}_{1}$ and $d^{[4]}_{2}$ can be written as
\begin{align}
d^{[4]}_{1} 
 &= [\delta^{[4]},S] \comma \quad
d^{[4]}_{2}
 = [Q^{[4]}_{0},S] \comma
\end{align}
 where $S$, carrying  degree $2$ under the current grading,
 is given by
\begin{align}
S &= {1\over2}\theta_{pq}F_{\ptil\qtil} \comma \\
F_{\ptil\qtil} 
&= -(\theta_{\ptil}\del\theta_{\qtil} 
       - \theta_{\qtil}\del\theta_{\ptil})\theta_{+}
   + 2i\lambda_{+}^{-1}\theta_{+}
       (\lambda_{\ptil}\delx^{-}_{\qtil} - \lambda_{\qtil}\delx^{-}_{\ptil}) 
 \period
\end{align}
Furthermore, it is easy to confirm that the double commutators
 all vanish:
\begin{align}
\bigl[[\delta^{[4]},S\bigr],S]
 &= \bigl[[Q^{[4]}_{0},S],S\bigr]
  = 0 \period
\end{align}
Combining these results, the similarity transformation 
(\ref{Stransf}) is proved. 

To summarize, we have demonstrated that 
 the BRST operator $\Qhat$ for the EPS formalism can be 
mapped to  $\Qbar$, the one  for 
 the light-cone GS formalism,  by a series of similarity transformations
 in the manner
\begin{align}
\Qbar
 &=e^{c_{0}K}e^{T}e^{S}e^{R}e^{Z}e^{Y}e^{X}
    \Qhat
   e^{-X}e^{-Y}e^{-Z}e^{-R}e^{-S}e^{-T}e^{-c_{0}K} \period
\end{align}
%%%%%%%%%%%%%%%%%%%
\subsection{Zero Modes and Cohomology }
%%%%%%%%%%%%%%%%%%
Having reduced $\Qhat$ to $\Qbar=\delta_{lc}+Q_{lc}$,
 we now discuss in some detail the 
 cohomology of the latter operator, focusing in particular on 
  the zero mode sector. 

Let us write the general state of our system as $\ket{\Phi} \otimes 
\ket{\Psi} $, where $\ket{\Phi} $ consists of the light-cone 
fields $p^\pm, \al^i_n, S^a_n$ and the ghost zero mode $c_0$, 
while $\ket{\Psi} $ is composed of the remaining fields. 
First, states with non-zero mode excitations
 in $\ket{\Psi} $ are unphysical by the standard argument:
 As discussed previously,  $\acom{\Qbar}{K} = 
\acom{\delta_{lc}}{K} =N$ holds with $N$ being the level-counting operator
 for the $\Psi$-modes, and non-trivial cohomology of $\Qbar$ occurs 
 only in the sector where $N=0$ \ie without non-zero $\Psi$ modes. 
We shall denote such sector as $\ket{\Phi} \otimes \ket{\Psi_0} $, where 
the subscript $0$ signifies that only the zero 
 modes are present. 

Now in this sector, $\Qbar$ is effectively reduced to $\Qbar_0 \equiv 
\delta_{lc,0} +  Q_{lc}$, where $\delta_{lc,0}$ is the zero mode part
 of $\delta_{lc}$  given by 
\begin{eqnarray}
\delta_{lc,0} &=& \lam^\adot_0 p_{\adot,0} -2 \lam^a_0 \Stil^{-a}_0
\period
\end{eqnarray}
This operator serves as the nilpotent operator with respect to which 
$(\lam^\adot_0, \omega_{\adot,0}, \theta^\adot_0, p_{\adot,0})$ and 
$(\lam^a_0, \omega_{a,0}, \Stil^{+a}_0, \Stil^{-a}_0)$ form quartets. To make
 the discussion transparent, let us denote each of these zero-mode quartets 
generically  as $(\ga, \be, c, b)$,
 the members of which 
 satisfy the (anti-)commutation relations\footnote{For instance,
 $(b,c)$ ghosts  here stand for $(\theta^\adot, p_{\adot,0})$ or
 $(\Stil^{+a}_0, \Stil^{-a}_0)$. Although the symbol $(b,c)$ have already 
 been used to denote $(b_5, c_\fivetil)$ pair, we believe there will be 
 no confusion.}
\begin{eqnarray}
\com{\ga}{\be} &=& 1\comma \qquad \acom{c}{b} = 1\period
\end{eqnarray}
In the  $\ga$-$c$ representation of the 
states, 
the role of $\delta_{lc,0}$ is played by the operator
\begin{eqnarray}
q &\equiv & \ga b = \ga{\del \over \del c} \comma 
\end{eqnarray}
and we will first study the cohomology of this nilpotent ``BRST'' operator. 
This depends on the choice of the Hilbert space. Most notably, if we allow 
 a special operator of the form $\xi \equiv \ga^{-1} c$, then since 
$\acom{q}{\xi} =1$ the cohomology becomes trivial. Indeed  any 
 $q$-closed state $\psi$ can be written as $\psi = \acom{q}{\xi} \psi 
 = q \xi \psi + \xi q \psi = q (\xi \psi)$, which is $q$-exact. 
Thus we need to exclude $\xi$ from our Hilbert space. Actually, 
 in the case of $q=\lam^\adot_0 p_{\adot,0}$, exclusion is automatic
 if we demand  $SO(8)$ covariance. However, as our similarity transformation 
 only respects $U(4)$ covariance in the intermediate steps, it is safe 
 and do no harm to exclude such an operator explicitly. This however 
 does not mean that  other operators with $\ga^{-1}$ 
 factor need be excluded,
 since they do not trivialize the cohomology as $\xi$ does.
 In fact we  need such  operators, for instance $Y=-2i \lampinv c_\ptil
 d_p $ in (\ref{defY}), 
 in the similarity transformation.
Now for $\ga^{-1}$ factor to be admissible, we need to exclude 
states which are localized at $\ga=0$ namely ones with $\delta(\ga)$ 
factor, on which $\ga^{-1}$ is ill-defined. 

With the above specification of the Hilbert space, let us analyze 
 the cohomology of $q$. 
The most general allowed wave function is of the  form
\begin{eqnarray}
\psi &=& f(\ga) + \ftil(\ga) c\comma 
\end{eqnarray}
where $f(\ga)$ and $\ftil(\ga)$ are arbitrary functions not localized 
 at $\ga=0$ and in addition
 $\ftil(\ga)$ should not be proportional to $\ga^{-1}$ to avoid the 
 structure $\ga^{-1}c$.
 Then $q$-closedness condition reads $q\psi = \ga \ftil(\ga)
=0$. Since we exclude the solution $\ftil \sim \delta(\ga)$, this demands
$\ftil(\ga)=0$ and $\psi = f(\ga)$. On the other hand the most general 
 $q$-exact state is of the form $q\chi$ with 
$\chi = g(\ga) + \gtil(\ga) c$, 
where the same restrictions as for $f,\ftil$ apply to $g$ and $\gtil$. 
Since $q\chi = \ga\gtil(\ga)$, 
$q$-closed $\psi$ is $q$-exact if $f(\ga)$ can be written as $\ga \gtil(\ga)$. 
Since $\gtil(\ga)\ne \ga^{-1}$ this means that all $q$-closed $\psi$, except 
$\psi =1$, are trivial. Hence we find that the cohomology 
 of $q$ in our Hilbert space consists solely of the vacuum state annihilated
 by $\be$ and $b$. 

Now we apply this result to our original problem and study the cohomology 
 of $\Qbar$. $\Qbar$-closed state of the form $\ket{\Phi} \otimes \ket{\Psi_0}
 $ must satisfy $Q_{lc} \ket{\Phi} =0$ and $\delta_{lc} \ket{\Psi_0} =0$. 
If $\ket{\Psi_0} $ is not the vacuum state $\ket{0} $, it can be
written as $\delta_{lc,0} \ket{\Xi_0} $. But then, 
$\ket{\Phi} \otimes \ket{\Psi_0} = (Q_{lc} + \delta_{lc,0})
 \ket{\Phi} \otimes \ket{\Xi_0} $ and such a state is cohomologically trivial. 
Hence the cohomology of $\Qbar$ is reduced to that of $Q_{lc}$ in 
 the space $\ket{\Phi} \otimes \ket{0} $. The rest of the analysis is 
entirely  parallel to the bosonic case and we reproduce
 the light-cone on-shell spectrum. 
%%%%%%%%%%%%%%%%%%%%%
\section{Discussions}
%%%%%%%%%%%%%%%%%%%%%%%%
In this article, making use of the systematic technique developed
 in our previous work \cite{AK2}, we have been able to construct 
 an explicit quantum mapping between the light-cone quantized GS string 
 and the extended version of the pure spinor formalism in the form of 
 a similarity transformation. As already remarked in the introduction, 
 this allows one to see the mechanism of the decoupling of unphysical 
 degrees of freedom in a transparent way, which is an improvement over 
 the earlier work \cite{Berk0006} requiring infinite number of ghosts. 

There are of course many further works to be performed. 
An immediate question of interest is to study how some basic operators 
 of the LCGS are represented in the EPS and vice versa. For instance, 
it would be intriguing to see the counterparts of the physical oscillators 
 $\al^i_n$ and $S^a_n$ on the EPS side, which are the analogues 
 of the DDF operators. They are expected to be closely  related to 
the massless vertex operators already constructed in \cite{AK1}. 
Another obvious project is to understand 
how the supercharges of the two formulations are
 mapped into each other. This may not be quite straightforward since 
 the supercharges in the LCGS act only on the restricted physical space
 satisfying the on-shell condition $L_0^{lc} \ket{\phi} =0$, whereas the ones 
 in the EPS act on all the components of $x^\mu$ and $\theta^\al$ including
 the unphysical ones. In the BRST framework, this means that we must 
 compare the action of the supercharges and their closure relations 
 at the cohomological level, \ie up to BRST-exact terms. Once 
 these correspondences are clarified, it will become possible to map 
 the amplitude calculation as well. 
This would clarify  the rules of computations
 on the EPS side, which have not yet been spelled out. 

More challenging task is the extraction of information 
 on the underlying local symmetry of our extended theory, crucial 
 for the eventual construction of the fundamental action.
 As we have identified  the structure of all the quartets in this work,
 this knowledge should give some useful hints. In this regard, let us 
 recall that the reparametrization $(b,c)$ ghosts appeared  as  part of the 
ghosts $(b_P, c_\Ptil)$ which are introduced to remove the PS constraints. 
This suggests that in this type of formulation bosonic and fermionic 
 local symmetries are intimately intertwined. Clarification of this 
 structure  certainly deserves further investigation. 

 The problems listed 
 above and more are currently under study  and we hope to report our 
 progress elsewhere. 
 
%%%%%%%%%%%%%%%%%%%%%%%
\par\bigskip\noindent
{\large\bf Acknowledgment}\par\smallskip\noindent
%%%%%%%%%%%%%%%%%%%%%%%%%%%%%%%%%%%%
Y.~K. would like to thank M.~Kato 
 for a useful discussion. 
The research of Y.K. is supported in part by the 
 Grant-in-Aid for Scientific Research (B) 
No.~12440060 and (C) No.~15540256 from the Japan 
 Ministry of Education,  Science and Culture. 
%%%%%%%%%%%%%%%%%%%%%%%%%%%%%%%%%%%%%%%%%%%%%%%%%%%%%%%%%%%%%%%%
\newpage
\appendix
%%%%%%%%%%%%%%%%%%%%%%%%%%%%%%%%%%%%%%%%%%%%%%%%%%%%%%%%%%%%%%%%
%%%%%%%%%%%%%%%%%%%%%%%%%%%%%%%%%%%%%%%%%%%%%%%%%%%%%%%%%%%%%%%%
\setcounter{equation}{0}
\renewcommand{\theequation}{A.\arabic{equation}}
\section*{Appendix A: \ \
Conventions and Useful Formulas}
%%%%%%%%%%%%%%%%%%%%%%%%%%%%%%%%%%%%%%%%%%%%%%%%%%%%%%%%%%%%%%%%
In this appendix,  our conventions and some 
 useful formulas are collected. 
%%%%%%%%%%%%%%%%%%%%%%%%%%%%%%%%%%%%
\subsection*{A.1.\ \ Spinors and $\Ga$-matrices in real basis}
%%%%%%%%%%%%%%%%%%%%%%%%%%%%%%%%%%%%%
$32\times 32$ $SO(9,1)$ Gamma matrices are denoted by $\Ga^\mu, 
 (\mu=0,1,\ldots, 9)$ and obey the Clifford algebra 
 $\acom{\Ga^\mu}{\Ga^\nu} =2\eta^{\mu\nu}$. Our metric convention is
 $\eta^{\mu\nu} = (-,+,+,\ldots, +)$. The 10-dimensional chirality 
 operator is taken to be $\Gabar_{10}=-\Ga^0\Ga^1\cdots \Ga^9$ and 
 it satisfies  $\Gabar_{10}^2=1$. 

In the Majorana or real basis (R-basis
 for short), $\Ga^\mu$ are all real and unitary. Within the R-basis, 
 we define the Weyl basis to be the one in which $\Gabar_{10} =${\rm diag}\
 $(1_{16}, -1_{16})$, where $1_{16}$ is the $16\times 16$ unit matrix. 
In this basis, a general 32-component spinor $\Lam$ is written as 
$\Lam = \vecii{\lam^\al}{\lam_{\al}} $, where $\lam^\al$ 
and $\lam_\al$ are chiral and anti-chiral respectively, 
with $\al=1 \sim 16$. Correspondingly, $\Ga^\mu$, which flips chirality, 
 takes the structure 
\begin{eqnarray}
\Ga^\mu &=& \matrixii{0}{(\ga^\mu)^{\al\be}}{(\ga^\mu)_{\al\be}}{0} 
\comma \label{GainWeyl} 
\end{eqnarray}
where the $16\times 16$ $\ga$-matrices $(\ga^\mu)^{\al\be}$ and 
 $(\ga^\mu)_{\al\be}$ are real symmetric and satisfy 
\begin{eqnarray}
(\ga^\mu)_{\al\be} (\ga^\nu)^{\be\ga} 
+ (\ga^\nu)_{\al\be} (\ga^\mu)^{\be\ga} =2\eta^{\mu\nu}\delta_\al^\ga 
\period \label{gaClif}
\end{eqnarray}
For $\ga^m, m=1\sim 9$, we have $(\ga^m)^{\al\be} = (\ga^m)_{\al\be}$. 
 As for $\ga^0$, we will use the convention $(\ga^0)_{\al\be} 
= -(\ga^0)^{\al\be} = \delta_{\al\be}$. 
%%%%%%%%%%%%%%%%%%%%%%%%%%%%%%%%%%%%%%%%%%%%%%%
\subsection*{A.2.\ \ {\boldmath $U(5)$} basis}
%%%%%%%%%%%%%%%%%%%%%%%%%%%%%%%%%%
The spinor representations for $SO(9,1)$ 
 and $SO(10)$ can be conveniently 
 constructed with the use of 5 pairs of 
 fermionic oscillators $(b_P, b_P^\dagger)_{P=1\sim 5}$
 satisfying the anti-commutation
 relations $\acom{b_P}{b_Q^\dagger} = \delta_{PQ}$. States are built upon 
 the oscillator vacuum, to be  denoted by $\ket{+} $, annihilated by 
all the $b_P$'s. 
 $b_P$ and $b_P^\dagger$ transform 
 respectively as $5$ and $\bar{5}$ of $U(5)$ subgroup\footnote{
For $SO(9,1)$, it is a Wick rotation of $U(5)$ but we continue to 
 use this terminology following common usage.}. 
$\Ga^\mu$ matrices can then be regarded as linear operators in this 
 Fock space  and in the case of $SO(9,1)$ they are identified as 
\begin{eqnarray}
\Ga^{2p} &=& {1\over i} (b_p-b_p^\dagger)\comma 
 \quad \Ga^{2P-1} = b_P +b_P^\dagger \comma \label{Gab1} \\
\Ga^0 &=& -(b_5-b_5^\dagger) \comma \label{Gab2}
\end{eqnarray}
Here and throughout, we use the notation 
$P = (p, 5)$, where the lower case letter $p$ runs from 1 to 4. 

The states built upon $\ket{+} $ and their conjugates are defined  as 
\begin{eqnarray}
\ket{P_1P_2 \ldots P_k} &\equiv & b_{P_1}^\dagger \cdots 
 b_{P_k}^\dagger \ket{+} \comma  \qquad 
\bra{P_1P_2 \ldots P_k} \equiv  \bra{+} b_{P_k} b_{P_{k-1}}
 \cdots b_{P_1} \period \label{U5ketbra} 
\end{eqnarray}
Further, we define
\begin{eqnarray}
\ket{-} &\equiv & b_1^\dagger b_2^\dagger \ldots b_5^\dagger \ket{+}
 = {1\over 5!} \ep_{P_1P_2 \ldots P_5} b_{P_1}^\dagger b_{P_2}^\dagger
 \cdots b_{P_5}^\dagger \ket{+} \comma \label{minusket} \\
\ket{\Ptil_1 \ldots \Ptil_k} &\equiv & {1\over (5-k)!} 
 \ep_{P_1 \ldots P_k Q_{k+1} \ldots Q_5} \ket{Q_{k+1} \ldots Q_5} 
 \comma \label{tildeket}
\end{eqnarray}
and their corresponding conjugates, where $\ep_{12345} \equiv 1$. 
These states satisfy the orthonormality relations 
\begin{eqnarray}
\bk{+}{+} &=& \bk{-}{-} =1 \comma \\
 \bk{P_1 \ldots P_k}{Q_1 \ldots Q_k} &=& 
\bk{\Ptil_1 \ldots \Ptil_k}{\Qtil_1 \ldots \Qtil_k}
 = \delta^{P_1 \ldots P_k}_{Q_1\ldots Q_k} \period 
\end{eqnarray}
In this basis, chiral and anti-chiral spinors can be written as 
\begin{eqnarray}
 \mbox{chiral:} \qquad \ket{\lam} 
 &=& \lam_+\ket{+} + \half \lam_{PQ} \ket{PQ} + \lam_\Ptil \ket{\Ptil} 
 \comma \label{U5chiral} \\
 \mbox{anti-chiral:} \qquad \ket{\psi} 
 &=& \psi_-\ket{-} + \half \psi_{\Ptil \Qtil} \ket{\Ptil \Qtil} 
+ \psi_P \ket{P} 
 \period \label{U5achiral}
\end{eqnarray}
The charge conjugation matrix in this basis is given 
 by
\begin{eqnarray}
C &=& -(b_5-b_5^\dagger) (b_1-b_1^\dagger)
 \cdots (b_4-b_4^\dagger) = \Ga^0 \Ga^2\Ga^4 \Ga^6\Ga^8 \comma 
\label{U5C}
\end{eqnarray}
and satisfies $C^2=-1$. 
Its action on the states is  
\begin{eqnarray}
C\ket{+} &=& \ket{-} \comma \quad C\ket{PQ} = -\ket{\Ptil \Qtil}
 \comma \quad C \ket{\Ptil} = \ket{P} \comma \\
C\ket{-} &=& -\ket{+} \comma \quad C\ket{\Ptil \Qtil} = \ket{PQ} 
\comma \quad C\ket{P} = -\ket{\Ptil} \period \label{Caction}
\end{eqnarray}
%%%%%%%%%%%%%%%%%%%%%%%%%%%%%%%%%%
\subsection*{A.3.\ \ {\boldmath $SO(8)$} parametrization}
%%%%%%%%%%%%%%%%%%%%%%%%%%%%%%%%%
In the following, $\lam^\al$ and $\chi_\al$ denote 10D chiral
 and anti-chiral spinors respectively. 
$16\times 16$ $SO(8)$ chiral projectors  are defined by 
\begin{eqnarray}
P^\pm_8 &=& \half (1\pm \ga^9) \comma \qquad \ga^9 =
\ga^1 \ga^2 \cdots \ga^8\comma \qquad (\ga^9)^2=1 \period
\end{eqnarray}
In the R-basis, we may take $\ga^9$ to be  diagonal  \ie  $\ga^9 =
 {\rm diag}\ (1_8, -1_8)$. Then, 
 the decomposition of $\chi^\al$ and $\psi_\al$ is simply given by 
\begin{eqnarray}
\lam^\al &=& \vecii{\lam^a}{\lam^\adot} \comma \qquad 
\chi_\al =\vecii{\chi_a}{\chi_\adot}  \period
\end{eqnarray}
In this representation, $\ga^i\ (i=1\sim 8)$ can be written as
\begin{eqnarray}
\ga^i &=& \matrixii{0}{\ga^i_{a\adot}}{\ga^i_{\adot a}}{0} \comma 
\end{eqnarray}
where $\ga^i_{a\adot}$ are symmetric $SO(8)$ Pauli matrices 
 satisfying the anti-commutation relation and the Fierz identity:
\begin{eqnarray}
&&\ga^i_{a\adot} \ga^j_{\adot b} + \ga^j_{a\adot} \ga^i_{\adot b} 
 = 2\delta^{ij}\delta_{ab} \comma \\
&& \ga^i_{a\adot} \ga^i_{b\bdot} + \ga^i_{a\bdot}\ga^i_{\adot b} = 2\delta_{ab}
 \delta_{\adot\bdot} \period
\end{eqnarray}

$P^\pm_8$ are essentially $\ga^\pm$, which are the off-diagonal 
elements of $\Ga^\pm = {1\over \sqrt{2}} (\Ga^0 \pm \Ga^9)$. 
Precise relations are  
\begin{eqnarray}
 (P_8^\pm)_{\al\be} &=& \ovsqtwo \ga^\pm_{\al\be}
 \comma \qquad 
(P^\mp_8)^{\al\be}  = -\ovsqtwo \ga^{\pm \al\be} \period
\end{eqnarray}
In the R-basis, $(P^+_8)_{ab} = \delta_{ab}$ and $(P^-_8)_{\adot\bdot}
 =\delta_{\adot\bdot}$, so that raising and lowering of 
 chiral and anti-chiral indices  $a$ and $\adot$ are trivial. 
However, in the $U(5)$ basis, we must use $\pm \ga^\pm/\sqrt{2}$ 
for this purpose, the action of which is non-trivial. 

For the construction of the similarity transformation for 
 the superstring case described in Sec.~4, we often need 
 the expression of $SO(8)$ spinors in terms of $U(5)$ components. 
It is convenient to introduce the following abbreviated ket notations:
\begin{eqnarray}
\ket{\lam} &\equiv & \ket{\lam^a} \comma \qquad \ket{\lambar} \equiv 
 \ket{\lam^\adot} \comma \\
\ket{\chi} &\equiv & \ket{\chi_a} \comma \qquad \ket{\chibar}
 \equiv \ket{\chi_\adot} \period
\end{eqnarray} 
Then, their $U(4)$ decompositions are given by 
\begin{eqnarray}
\ket{\lam} &=& \lamp \ket{+} + \half \lam_{pq} \ket{pq} + \lam_\fivetil
\ket{\fivetil} \comma \\
\ket{\lambar} &=& \lam_{p5}\ket{p5} + \lam_\ptil\ket{\ptil}  \comma \\
\ket{\chi} &=& \chi_-\ket{-} + \half \chi_{\ptil\qtil}\ket{\ptil\qtil}
+\chi_5 \ket{5} \comma \\
\ket{\chibar} &=& \chi_q\ket{q} + \chi_{\qtil \fivetil} \ket{\qtil \fivetil}
\period
\end{eqnarray}
Together with the expression of $\Ga$-matrices in terms of 
 the fermionic oscillators $b_P, b^\dagger _P$ given in 
(\ref{Gab1}) and (\ref{Gab2}), one can convert 
various spinor bilinears from $SO(8)$ parametrization to $U(5)$ 
parametrizations. We just give one such example for illustration. 
Let $\lam^a$ and $\theta^a$ be 10D chiral $SO(8)$ chiral spinors and 
 consider the bilinear $\lam^a \theta_a$ in the R-basis. 
 As already remarked, in going to other basis we must interpret 
 this as $\lam^a (P^+_8)_{ab}\theta^b =\lam \ga^+ \theta/\sqrt{2} $. 
To compute this in  $U(5)$ basis, we should write 
this  as $ \bra{\lam} C\Ga^+ \ket{\theta} /\sqrt{2}$ 
 and use the identification $\Ga^+ = \sqrt{2}\, b_5^\dagger$. 
Then, using the known action of $b_5^\dagger$ and $C$ on $U(5)$ states,
 we  easily obtain the result 
$\lam^a \theta_a = -\lamp \theta_\fivetil -\lam_\fivetil \theta_+
 + (1/4) \ep_{pqrs5} \lam_q{rs} \theta_{pq}$. 

%%%%%%%%%%%%%%%%%%%%%%%%%%%%%%%%%%%%%%%%%%%%%%%%%%%%%%%%%%%%%%%%
\setcounter{equation}{0}
\renewcommand{\theequation}{B.\arabic{equation}}
\section*{Appendix B: \ \ Some Details of the Construction of Similarity 
 Transformations} 
In this appendix, we supply some further details of the 
construction of the similarity  transformations. 
%%%%%%%%%%%%%%%%%%%%%%%%%%%%%%%%%%%%%%%%%%%%%%%%%%%%%%%%%%%%%%%%
\subsection*{B.1.\ \ Bosonic string }
%%%%%%%%%%%%%%%%%%%%%%%%%%%%%%%%%%%%%%%%%%%%%%%%%%%%%%%%%%%%%%%%
Below we shall complete the proof of the similarity transformation 
(\ref{QQtil}), reorganized according to the degree as 
\begin{eqnarray}
Q &=& \delta + Q_0 + d_1+d_2+d_3= e^{-R} (\delta + Q_0) e^{R} 
 \nn\\
&=& \delta + Q_0 + \com{\delta}{R} + \com{Q_0}{R}
 + \half \com{\com{\delta}{R}}{R} + \half \com{\com{Q_0}{R}}{R} +\cdots \nn\\
&=& \delta + Q_0 +  \com{\delta}{R_2} + \left(\com{\delta}{R_3} +
\com{Q_0}{R_2} \right) + \left( \com{Q_0}{R_3} + 
\half \com{\com{\delta}{R_2}}{R_2} \right) + \cdots
\period  \label{apsimtrone}\nn \\
\end{eqnarray}
 $R$ is  given, as in (\ref{defR}), by 
\begin{eqnarray}
R &=& R_2 + R_3\comma \qquad R_2= \acom{\Khat}{d_1} \comma 
\qquad R_3 =\half \acom{\Khat}{d_2} \comma  \label{apdefR}
\end{eqnarray}
which  has been determined to satisfy $\com{\delta}{R_2+R_3} 
 =d_1+d_2$. Throughout,  our strategy is to make use of the 
 relations (\ref{Emtwo}) $\sim$ (\ref{Esix}) which follow from 
the nilpotency of $Q$ as well as the representations  
of $R$ given in (\ref{apdefR}) and appropriate graded Jacobi identities to circumvent
 explicit computations involving the rather complicated operator $d_1$. 

To verify (\ref{apsimtrone}) up to degree 2, we need to
 show that $\com{Q_0}{R_2} $ vanishes. Using (\ref{apdefR}) 
it can be computed as 
\begin{eqnarray}
\com{Q_0}{R_2} &=& \com{Q_0}{\acom{\Khat}{d_1} } 
=\com{\acom{Q_0}{\Khat} }{d_1}
 -\com{\Khat}{\acom{Q_0}{d_1} }  \period
\end{eqnarray}
The last term on the RHS vanishes by the first 
equation in  (\ref{eqE1}), while it is easy to check explicitly that 
 $\acom{Q_0}{\Khat} =0$. Thus we get $\com{Q_0}{R_2} =0$. 

Next we move on to degree 3. First, by using $R_2= \acom{\Khat}{d_1} $
 and graded Jacobi identities,  the double commutator 
$ \com{\com{\delta}{R_2}}{R_2} $ can easily be rewritten as 
$ \com{\com{\delta}{R_2}}{R_2} = -\half \com{\Khat}{\acom{d_1}{d_1} } $.
Applying the relation $(E_2)$ in (\ref{Etwo}), we may write this as 
\begin{eqnarray}
\com{\com{\delta}{R_2} }{R_2} &=&\com{\Khat}{\acom{\delta}{d_3} } + \com{\Khat}{\acom{Q_0}{d_2} }   \period
\end{eqnarray}
The first term can be directly computed to give $2d_3$, while the second 
 term, by using a Jacobi identity and the relation $\acom{Q_0}{\Khat}
 =0$,  can be shown to equal $-2\com{Q_0}{R_3} $. Combining these 
 results, we get 
\begin{eqnarray}
\half \com{\com{\delta}{R_2} }{R_2} + \com{Q_0}{R_3} &=& d_3 \comma 
\label{deltaRR}
\end{eqnarray}
which proves the validity of the similarity transformation 
at degree 3. The remaining task is to show that terms
with  higher degrees all vanish.

 For this purpose it is convenient to derive some 
 relations at degree 3 and 4. First from the structure of $Q_0$ and $d_3$ 
it is easy to see that $\acom{Q_0}{d_3} =0$. Due to the relation 
$(E_3)$ shown in (\ref{Ethree}), this in turn dictates $\acom{d_2}{d_1} =0$. 
Similarly, $\acom{d_2}{d_2} =0$ by inspection and this leads via 
the relation $(E_4)$ in (\ref{Efour}) to $\acom{d_3}{d_1} =0$. 
This then leads to $\com{R_3}{d_1} =0$ since, apart form a $b_0$ 
 factor, $R_3$ is actually proportional to $d_3$. 
Other useful vanishing relations at these degrees are:
 $\acom{\Khat}{R_2} =0$, which follows from the nilpotency of $\Khat$, 
and $\acom{\Khat}{d_3} = \acom{\Khat}{R_3} =0$, which can be checked
 explicitly.

Now we are ready to  show that 
the double commutators $\com{\com{\delta}{R_i} }{R_j} $ and 
$\com{\com{Q_0}{R_i} }{R_j} $ with $i=2,3$, except 
$\com{\com{\delta}{R_2} }{R_2} $, and the triple commutators 
$\com{\com{\com{\delta}{R_2} }{R_2} }{R_i} $ vanish. 
First, using $\acom{Q_0}{d_3} =0$, one  easily find
 that $\com{Q_0}{R_3} \propto L_0^{tot} R_3 $ and hence 
$\com{\com{Q_0}{R_3} }{R_3} =0$. Since we already know $\com{Q_0}{R_2} $
vanishes,  this shows that $\com{\com{Q_0}{R_i} }{R_j} $ all vanish. 
Next we can  prove a 
somewhat non-trivial relation  $\com{d_2}{R_2} = 0$ by rewriting 
$\com{d_2}{R_2} $ via a Jacobi identity as $2 \com{R_3}{d_1} -\com{\Khat}{
\acom{d_2}{d_1} } $. Since $d_2=\com{\delta}{R_3} $, this 
 is nothing but the vanishing of the double commutator 
  $\com{\com{\delta}{R_3} }{R_2}  =0$. This in turn implies 
$\com{\com{\delta}{R_2} }{R_3}  =0$ with the aid of a Jacobi identity, 
 since $\com{R_3}{R_2} = -\half \acom{\com{\Khat}{R_2} }{d_2} 
-\half \acom{\Khat}{ \com{d_2}{R_2} } =0$. Also, 
 $\com{\com{\delta}{R_3} }{R_3}  =\com{d_2}{R_3} =0$ is seen to hold
 by inspection. Finally, by using (\ref{deltaRR}) and $\com{d_3}{R_2} =0$
 which follows from previously derived vanishing relations, 
we find that the triple commutators
 $\com{\com{\com{\delta}{R_2} }{R_2} }{R_i} $  vanish. 
Thus, the similarity transformation terminates at  degree 3 and 
our assertion is proved.

%%%%%%%%%%%%%%%%%%%%%%%%%%%%%%%
\subsection*{B.2.\ \ Superstring}
%%%%%%%%%%%%%%%%%%%%%%%%%%%%%%%%
In this appendix, we shall supply some details of the construction 
 of the operator $R$ sketched in subsection 4.2.. 
To facilitate the discussion, we shall denote 
 the graded commutator $[A,B\}$ of two integrated operators by $AB$ 
 and suppress the superscript ``$[3]$'' on $\delta^{[3]}$, $Q^{[3]}_{0}$
 and $d^{[3]}_{n}$. In this notation, 
 the equations (\ref{eq:simQ3toQ4deg1}) $\sim$ (\ref{eq:simQ3toQ4deg5}) take the form 
\begin{align}
&\quad(\Etil_1):\quad d_{1} = \delta R_{2} \comma \\
&\quad(\Etil_2):\quad d_{2} = \delta R_{3} + Q_{0} R_{2} \comma \\
&\quad(\Etil_3):\quad d_{3} = \delta R_{4} + Q_{0} R_{3} 
 + {1\over2}(\delta R_{2})R_{2} \comma \\
&\quad(\Etil_4):\quad
 d_{4} = \delta R_{5} + Q_{0} R_{4} 
  + {1\over2}(\delta R_{2})R_{3} + {1\over2}(\delta R_{3})R_{2} 
  + {1\over2}(Q_{0} R_{2})R_{2} \comma
\\
&\quad(\Etil_5):\quad 
 0 = \delta R_{6} + Q_{0} R_{5} + \cdots \period
\end{align}
Also it will be useful to display the explicit equations which follow 
 from the nilpotency of $Q^{[3]}$. They read 
\begin{align}
&\quad(N_0):\quad  0 = 2\delta d_{1} + (Q_{0})^{2}, \\
&\quad(N_1):\quad  0 = \delta d_{2} + Q_{0}d_{1}, \\
&\quad(N_2):\quad  0 = \delta d_{3} + Q_{0}d_{2}, \\
&\quad(N_3):\quad  0 = \delta d_{4} + Q_{0}d_{3}, \\
&\quad(N_4):\quad  0 = Q_{0}d_{4}.
\end{align}
As already mentioned in the text, 
 the fact that $F$'s and $B$'s (anti-)commute 
 among themselves will be tacitly used throughout the subsequent 
 analysis.

Since we have already described the solution of $(\Etil_1)$ in the text, 
we begin with $(\Etil_2)$, which can be written as
$d_2-Q_0R_2 =\delta R_3$.  
To determine $R_{3}$, let us compute $d_{2}-Q_{0}R_{2}$ more concretely.
Substituting the explicit forms of $d_2, Q_0$ and $R_2$, 
 we obtain
\begin{align}
 d_{2} - Q_{0}R_{2}
 &= \theta_{p5}(B_{\ptil}+Q_{0}F_{\ptil}) 
    + \lambda_{p5}\lambda_{q5}F'_{\ptil\qtil} \period
\end{align}
Actually one can show that the first term on the RHS vanishes 
due to the nilpotency relation $(N_1)$:
Substituting $d_{1}=\lambda_{p5}F_{\ptil}$ 
 and $d_{2}=\theta_{p5}B_{\ptil} + \lambda_{p5}\lambda_{q5}F'_{\ptil\qtil}$,
$(N_1)$ reads  
\begin{align}
 0 &= \delta d_{2} + Q_{0}d_{1} = \lambda_{p5}(B_{\ptil}+Q_{0}F_{\ptil}) 
\period
\end{align}
Since $\lambda_{p5}$'s are algebraically independent, 
 $B_{\ptil}+Q_{0}F_{\ptil}$ itself must vanish. 
In this way the equation $d_{2} - Q_{0}R_{2} = \delta R_{3}$ above 
is reduced to $\lambda_{p5}\lambda_{q5}F'_{\ptil\qtil} = \delta R_{3}$. 
Now it can easily be solved for $R_{3}$ as
\begin{align}
R_{3} = \lambda_{p5}\theta_{q5}F'_{\ptil\qtil} \period
\end{align}

Next, we move on to the degree $3$ equation $(\Etil_3)$. 
From the structure of $R_{2}$ given in equation (\ref{defR2}), 
it is easy to see that 
 the double commutator term $\half (\delta R_{2})R_{2}$ vanishes. 
Hence,  $(\Etil_3)$ simplifies to 
\begin{align}
 d_{3} - Q_{0}R_{3} = \delta R_4   \period
\end{align}
where the explicit form of the LHS is worked out as 
\begin{align}
d_{3} - Q_{0}R_{3}
 &= \lambda_{p5}\theta_{q5}(B_{\ptil\qtil} + Q_{0}F'_{\ptil\qtil})
 + \lambda_{p5}\del\theta_{q5}\tilde{B}_{\ptil\qtil} \period
\label{d3Q0R3}
\end{align}
To see that this expression is actually $\delta$-exact, we need the 
 information from the nilpotency relation $(N_2)$. Upon substituting
$d_{2} = Q_{0}R_{2} + \delta R_{3}$, $(N_2)$ becomes 
\begin{align}
0= \delta d_{3} + Q_{0}d_{2}
 &= \lambda_{p5}\lambda_{q5}B_{\ptil\qtil} 
  + \lambda_{p5}\del\lambda_{q5}\tilde{B}_{\ptil\qtil}
  + \theta_{p5}\theta_{q5}(Q_{0}F'_{\ptil\qtil}) %XXX=0
  + \lambda_{p5}\lambda_{q5}(Q_{0}F'_{\ptil\qtil}) \\
 &= \lambda_{p5}\lambda_{q5}
   (B^{S}_{\ptil\qtil} + Q_{0}F'_{\ptil\qtil}
 - {1\over2}\del\tilde{B}^{S}_{\ptil\qtil})
 + {1\over2}(\lambda_{p5}\del\lambda_{q5} - \lambda_{q5}\del\lambda_{p5})
  \tilde{B}^{A}_{\ptil\qtil} \comma 
\end{align}
where the superscripts $S$ and $A$ stand for the symmetric and antisymmetric 
 parts, respectively. 
From this we immediately obtain 
$B^{S}_{\ptil\qtil} + Q_{0}F'_{\ptil\qtil}
 = {1\over2}\del\tilde{B}^{S}_{\ptil\qtil}$ and
$\tilde{B}^{A}_{\ptil\qtil} 
  = 0$. Using these relations  (\ref{d3Q0R3}) is reduced to
\begin{align}
d_{3} - Q_{0}R_{3}
 &= \lambda_{p5}\theta_{q5}(B^{A}_{\ptil\qtil} 
   + {1\over2}\del\tilde{B}^{S}_{\ptil\qtil})
   + \lambda_{p5}\del\theta_{q5}\tilde{B}^{S}_{\ptil\qtil} \period
\end{align}
It is now not difficult to see that the RHS can be written as $\delta R_4$, 
 where 
\begin{align}
R_{4} = {1\over2}\theta_{p5}\theta_{q5}B^{A}_{\ptil\qtil}
 + {1\over2}\theta_{p5}\del\theta_{q5}\tilde{B}^{S}_{\ptil\qtil} \period
\end{align}

In an entirely similar manner we can analyze $(\Etil_n)$ to obtain $R_{n+1}$
 for $n \ge 4$. For $n=4$, 
 the double commutator terms vanish  as in previous steps and 
$(\Etil_4)$ becomes  $d_4-Q_0R_4 = \delta R_5$. This time 
a simplifying feature sets in at this stage. 
Analysis of 
$d_{4}-Q_{0}R_{4}$ and the relation $0=\delta R_{4} + Q_{0}R_{3}$
 reveals that $d_{4}-Q_{0}R_{4}$ itself vanishes and hence 
 we get $ R_{5} = 0$. Above $n=5$, one can easily check that 
 $R_{n\ge 6}=0$ solve all the equations. This completes the construction 
 of $R$.

%%%%%%%%%%%%%%%%%%%%%%%%%%%%%
%\nullify{
%%%%%%%%%%%%%%%%%%%%%%%%%%%%%%%%%%

%}%end of nullify
%%%%%%%%%%%%%%%%%%%%%%%%%%%%%%%%%%%%%%
\end{document}